\documentclass[a4paper,11pt]{article}
\usepackage{pos}

\usepackage{tikz}
\usetikzlibrary{shapes.geometric, arrows}

\def\be{\begin{equation}\begin{aligned}}
\def\ee{\end{aligned}\end{equation}}
\def\nn{\nonumber}

\def\thefootnote{\arabic{footnote}}
\allowdisplaybreaks 
\numberwithin{equation}{section}

\title{Supergravity EFTs and swampland constraints}

\author*[a]{Niccol\`o Cribiori}
\author*[b,c]{Fotis Farakos}

\affiliation[a]{Max-Planck-Institut f\"ur Physik (Werner-Heisenberg-Institut), \\ 
   F\"ohringer Ring 6,  80805 M\"unchen, Germany}

\affiliation[b]{Dipartimento di Fisica e Astronomia ``Galileo Galilei''\\ Universit\`a di Padova, Via Marzolo 8, 35131 Padova, Italy}

\affiliation[c]{ INFN, Sezione di Padova \\
Via Marzolo 8, 35131 Padova, Italy}

\emailAdd{cribiori@mpp.mpg.de}
\emailAdd{fotios.farakos@pd.infn.it}

\abstract{In these proceedings, we review recent progress in analyzing the behavior of lower-dimensional supergravity theories when combined with swampland conjectures. 
We show that within supergravity the effectiveness and usefulness of swampland conjectures gets amplified, existing criteria can be intertwined and also new ones can be uncovered.
Furthermore, we elaborate on some previously unpublished work. This includes evidence for the possible existence of a novel conjecture on Yukawa couplings and an argument to constrain large classes of D-term inflationary models using known conjectures.
} 

\FullConference{%
  Corfu Summer Institute 2022 "School and Workshops on Elementary Particle Physics and Gravity",\\
  28 August - 1 October, 2022\\
  Corfu, Greece}

\tableofcontents

\begin{document}
\maketitle

\newpage

\section{Introduction and discussion}

One of the main lessons that has been learned from quantum gravity is that not any low energy effective theory (EFT) that looks consistent remains so when coupled to dynamical gravity. In fact, the vast majority of effective theories do not and, as such, they are said to be in the swampland. With our present understanding, the swampland program \cite{Vafa:2005ui,Palti:2019pca,Agmon:2022thq} is a web of conjectures and statements which capture fundamental aspects of quantum gravity and in turn constrain the form of low energy effective theories. Even if for technical reason the majority of work testing swampland conjectures is performed on supersymmetric models, among the most solid and far reaching conjectures are statements like the absence of global symmetries in quantum gravity or the fact that gravity has to be the weakest force. As such, these statements do not assume supersymmetry and in fact they may be independent from it. As of now, it is not clear whether any of the known swampland conjectures implies supersymmetry in some form and at some energy scale. Indeed, it would be extremely important to be able to argue for the existence of supersymmetry just from swampland reasoning. While we are not able to perform such a step, in these proceedings we collect known and new evidence for the fact that, for what concerns swampland conjectures, effective supergravity theories are somehow well behaved with respect to generic effective theories. A priori this had not to be the case, for there is indeed no evidence at present that swampland constraints imply (low energy) supersymmetry.

Let us stress it from the very beginning: by no means we are saying that supergravity theories are automatically outside the swampland. This cannot be right. Instead, we want to point out that supergravity is better behaved than general effective theories, with respect to swampland conjectures. In fact, we will argue in favor of this assertion by showing that certain constraints are automatically implemented in supergravity models without the need for almost any additional input, except for some (well-established) conjectures. It is then tempting to think that, for some reason unknown to us (which is perhaps supersymmetry itself), supergravity has built-in the property to link swampland conjectures to one another. 
Schematically, we mean that 
\begin{equation}
\nn
\text{SUGRA + conjecture A} \quad \Longleftrightarrow \quad \text{SUGRA + conjecture B}, 
\end{equation}
where $A$ and $B$ are here two different swampland conjectures. 
This phenomenon can have important consequence within the swampland program, since it can furnish a direct method of structuring and reinforcing the aforementioned web of swampland conjectures.
A bolder statement is that one could potentially use supergravity as tool to generate new conjectures, that is 
\begin{equation}
\nn
\text{SUGRA + known conjecture(s)} \quad \Longrightarrow \quad \text{SUGRA + new conjecture(s)}. 
\end{equation}
In this review, we will discuss various examples of both statements.

Supergravity is the theory of local supersymmetry. Since its discovery \cite{Freedman:1976xh,Deser:1976eh}, it has been widely employed in a variety of different areas in fundamental high energy physics in such a way that it is hard to think of an aspect in which supergravity has not been employed at all. Arguably, two of the main motivations why it is still alive after almost half a century are the fact that supergravity is the low energy limit of the superstring, and indeed it even captures strong coupling in its eleven-dimensional formulation, and the fact that supergravity can be related to (conformal) field theories through holography. Indeed, the vast majority of the works in string phenomenology and holography are using supergravity in one way or another, and sometimes even unconsciously.

However, supergravity is not string theory. As such, one should not take for granted that supergravity is well behaved for what concerns quantum gravity. Nevertheless, in some cases it is, and in these cases it is probably much better behaved that any other effective field theory one may come up with.

In this contribution, we give evidence that supergravity is indeed a very good starting point when studying low energy effective theories in a swampland perspective. We will mainly present two related but not completely equivalent approaches. Our guiding principle is purely bottom-up: we take swampland conjectures as principles of quantum gravity and then use them actively to constrain low energy effective field theories. In particular, we do not aim at testing swampland conjectures, rather we want to understand their consequences assuming them to be fundamental quantum gravity features.

First, we consider the standard supergravity action from textbooks and apply directly swampland conjectures. This approach is perhaps coarse, but we will argue that it leads to non-trivial results nevertheless. In particular, we will show that certain swampland conjectures, taken as they stand, can be automatically implemented within supergravity at the cost of introducing very mild assumptions, like some form of charge quantization. 
Second, we will review (not in a comprehensive manner though) a related approach in which supergravity is first supplemented by additional ingredients in the low energy, as strings or membranes, and then swampland conjectures are investigated in such an enriched setup. This approach is perhaps not completely bottom-up, for it draws information and inspiration from string theory models, but the analysis is ultimately four-dimensional. 
We believe that these two lines of research are primary examples of the a priory unexpected well-behaviour of supergravity with respect to swampland conjectures. 
Finally, we focus on further constraints which most clearly are inspired by the very structure of supergravity. We point out what could be a potentially new swampland conjecture involving Yukawa couplings and we analyze the consequences of the so called Festina Lente bound on D-term inflation, concluding that the majority of inflationary models of this type are generically in tension with it.
Throughout this review, we will mostly work in Planck units, but we will restore the Planck mass $M_P$ explicit in some relevant formulae.  Any misconception in the presentation of the work by other authors is of course ours.

Before entering the main topic of the discussion, we need to review basic elements of four-dimensional supergravity with four and eight supercharges. This is the topic of the next section.

\section{Basic elements of supergravity in four dimensions}

In this section, we review the ingredients of four-dimensional supergravity with four and eight supercharges that we are going to employ in what follows. For more details, we refer to standard textbooks, such as \cite{Wess:1992cp,Freedman:2012zz,DallAgata:2021uvl}.

\subsection{$\mathcal{N}=1$ supergravity}
\label{sec:N=1rev}

The minimal theory in four dimensions has four preserved supercharges and it is denoted as $\mathcal{N}=1$ supergravity. Due to the low amount of supersymmetry, it is the theory whose structure is constrained the less. On the one hand, this makes it an efficient and versatile tool for model building, on the other hand it may render the identification of the ultraviolet origin of a generic model rather obscure, in particular due to the lack of control over corrections.

The main ingredients of the theory are the gravity multiplet and matter multiplets, among which we recall chiral and vector multiplets. Other matter representations are known, such as linear multiplets, but we will not need them extensively in what follows. The gravity multiplet contains the graviton and the gravitino. Chiral multiplets are arguably the simplest matter multiplets and are made up of a complex scalar and a majorana (or Weyl) fermion, $\{z,\chi\}$. Vector multiplets, instead, contain one fermion and one real vector, $\{\lambda, A_\mu\}$, but no scalars. 
When looking at off-shell representations, chiral multiplets contain one complex scalar auxiliary fields while vector multiplets a real one.

The Lagrangian is completely fixed by three (or better two) functions of the scalar fields. These are: the K\"ahler potential $K(z,\bar z)$, the superpotential $W(z)$ and the gauge kinetic function $f(z)$. More precisely, $K$ and $W$ are not independent, for they are related by K\"ahler transformations, $K\to K+\omega(z)+\bar \omega(z)$ and $W\to e^{-\omega}W$, with $\omega=\omega(z)$ an arbitrary holomorphic function, but one can construct the K\"ahler invariant quantity
\begin{equation}
G=K+\log W \bar W,
\end{equation}
which is indeed the only combination of $K$ and $W$ appearing in the Lagrangian. The bosonic sector of the theory reads\footnote{With respect to \cite{Freedman:2012zz}, we are following conventions in which the gauge kinetic function is $(f)_{there} = -i\, (f)_{here}$, or equivalently $({\rm Re}f)_{there} = ({\rm Im}f)_{here}$, $({\rm Im}f)_{there} = -({\rm Re}f)_{here}$.}
\begin{equation}
\label{N=1LagBos}
\begin{aligned}
e^{-1}\mathcal{L} &=\frac12 R - g_{i\bar\jmath} D_\mu z^i D^\mu \bar z^{\bar \jmath} -\frac14 {\rm Im}f_{\Lambda \Sigma} F_{\mu\nu}^{\Lambda} F^{\mu\nu\,\Sigma} -\frac 18 {\rm Re}f_{\Lambda\Sigma} \epsilon^{\mu\nu\rho\sigma} F_{\mu\nu}^\Lambda  F_{\rho\sigma}^\Sigma -V,
\end{aligned}
\end{equation}
where the scalar potential is
\begin{equation}
V = e^K\left(g^{i\bar\jmath}D_{i}W\bar D_{\bar \jmath} \bar W-3 W \bar W\right) + \frac12 ({\rm Im }f^{-1})^{\Lambda \Sigma} \mathcal{P}_\Lambda \mathcal{P}_\Sigma.
\end{equation}
Here, the covariant derivative on the scalar fields is
\begin{equation}
\label{covderz}
D_\mu z^i = \partial_\mu z^i - A^\Lambda_\mu k_\Lambda^i ,
\end{equation}
while on the superpotential is
\begin{equation}
D_i W = \partial_i W + W \partial_i K .
\end{equation}
The indices $i,j=1,\dots,n_C$ are counting the number of chiral multiplets, while $\Lambda,\Sigma=1,\dots,n_V$ the number of vector multiplets.
Notice that we are already considering the gauged theory. In particular, we introduced the real moment maps $\mathcal{P}_\Lambda(z,\bar z)$ and the Killing vectors $k_\Lambda^i = - ig^{i\bar\jmath}\partial_{\bar \jmath}\mathcal{P}_\Lambda$, 
which correspond to the infinitesimal transformations 
\begin{equation}
\label{defkillv}
\delta z^i = \alpha^\Lambda k_\Lambda^i \,. 
\end{equation}
They are such that 
\begin{equation}
[k_\Lambda, k_\Sigma] = {f_{\Lambda \Sigma}}^\Gamma k_\Gamma,
\end{equation}
where ${f_{\Lambda \Sigma}}^\Gamma$ are the structure constants of the gauge group and have to satisfy the equivariance consistency condition 
\begin{equation}
k^i_\Lambda g_{i \bar \jmath}k^{\bar \jmath}_\Sigma-k^i_\Sigma g_{i \bar \jmath}k^{\bar \jmath}_\Lambda = i {f_{\Lambda \Sigma}}^\Gamma \mathcal{P}_\Gamma.
\end{equation}
Since Killing vectors are derivatives of the moment maps, two $\mathcal{P}_\Lambda$ differing by a constant vector $\xi_\Lambda \in \mathbb{R}^{n_V}$ give rise to the same $k_\Lambda$. This ambiguity in choosing the $\mathcal{P}_\Lambda$ up to $\xi_\Lambda$ gives rise to the existence of physical $\xi_\Lambda$-dependent couplings corresponding to Fayet-Iliopouls D-terms in the limit of global supersymmetry.

The only piece of information from the fermionic sector that we are going to employ extensively is the gravitino covariant derivative, which reads
\begin{equation}
D_\mu\psi_\nu = \left(\partial_\mu+\frac14 \omega_{\mu}^{ab}\gamma_{ab}+\frac i2 v_\mu\gamma_*-\frac i2 A_\mu^\Lambda \mathcal{P}_\Lambda\gamma_*\right)\psi_\nu \,. 
\end{equation}
Here, $v_\mu =-\frac i2 \left(\partial_\mu z^i \partial_i K-\partial_\mu\bar z^{\bar\jmath}\partial_{\bar\jmath}K\right)$ is a composite connection indicating that the manifold of the scalar fields is K\"ahler-Hodge.\footnote{A K\"ahler-Hodge manifold, or K\"ahler manifold of restricted type, is a K\"ahler manifold equipped with a complex line bundle $L$ such that the first Chern class is equivalent to the (de Rahm) cohomology class of the K\"ahler form, $c_1(L) = [\mathcal{K}]$. This line bundle can be equipped with a connection $\partial K$. Then, fermions are sections of the associated U$(1)$-bundle with connection $v = {\rm Im}(d\partial )$.}
Contrary to what happens for extended supergravity, especially for $\mathcal{N}>2$, in the $\mathcal{N}=1$ theory there is no restriction on the form of the scalar manifold, which can be any K\"ahler-Hodge variety. For completeness, we report also the supersymmetry transformations of the fermions, which read
\begin{align}
\delta P_L \psi_\mu&=D_\mu P_L \epsilon+\frac12 \gamma_\mu e^{\frac K2}W P_R \epsilon +\mathcal{O}(\psi,\chi,\lambda),\\
\delta \chi^i&=\frac{1}{\sqrt 2} \gamma^\mu D_\mu z^i \epsilon-\frac{1}{\sqrt 2}e^{\frac K2}g^{i \bar \jmath}\bar D_{\bar \jmath} \bar W \epsilon + \mathcal{O}(\psi,\chi,\lambda),\\
\delta \lambda^\Lambda &=\frac14 \gamma^{\mu\nu} F_{\mu\nu}^\Lambda\epsilon + \frac i2 \gamma_* ({\rm Im}f^{-1})^{\Lambda \Sigma} \mathcal{P}_\Sigma \epsilon + \mathcal{O}(\psi,\chi,\lambda).
\end{align}
The quantity
\begin{equation}
m_{3/2} = e^{\frac K2} W
\end{equation}
is the covariantly-holomorphic gravitino Lagrangian mass parameter. We will refer to it just as the gravitino mass in what follows, but one has to keep in mind that the concept of mass is strictly well-defined only in flat space.

The classical two-derivatives Lagrangian can be supplement with various corrections. Later on, we will discuss certain higher-derivative corrections to the Einstein--Hilbert term which in an appropriate normalization take the form \cite{Cecotti:1985mf,Cecotti:1987mr,Buchbinder:1988tj} 
\begin{equation}
\label{R2corrections}
-\frac{1}{96\pi}\int {\rm Im}\tilde f\, {\rm tr} \left(R\wedge * R\right) - \frac{1}{96\pi}\int {\rm Re} \tilde f \,{\rm tr}\left(R\wedge R\right)+\dots
\end{equation}
where dots contain $R_{\mu\nu}R^{\mu\nu}$ and $R^2$ terms which can be combined with $R\wedge *R$ to reconstruct the Gauss-Bonnet coupling $\frac{1}{192\pi} \int {\rm Im}\tilde f\, E_{GB} * 1$, with
\begin{equation}
E_{GB} = R_{\mu\nu\rho\sigma}R^{\mu\nu\rho\sigma} - 4 R_{\mu\nu} R^{\mu\nu} + R^2.
\end{equation}
The structure of the terms \eqref{R2corrections} resembles closely that of the kinetic and theta-term of the vector fields in the original Lagrangian \eqref{N=1LagBos}. Furthermore, we introduced an holomorphic function $\tilde f = \tilde f(z)$ parametrizing the coupling of the scalar fields to these corrections. Finally, we point out that the terms \eqref{R2corrections} arise from a superspace contribution
\begin{equation}
\frac{1}{24 \pi i}\int d^4 x d^2\Theta 2\mathcal{E} \tilde f(z)\, \mathbb{W}^2 + c.c.,
\end{equation}
where the composite chiral multiplet 
\begin{equation}
\mathbb{W}^2 = \mathcal{W}^{\alpha\beta\gamma}\mathcal{W}_{\alpha\beta\gamma} - \frac14 \left(\overline{\mathcal{D}}^2-8\mathcal{R} \right) \left(2\mathcal{R}\overline{\mathcal{R}} + \mathcal{G}^a \mathcal{G}_a\right)
\end{equation}
is constructed out of the chiral $\mathcal{W}_{\alpha\beta\gamma}$, $\mathcal{R}$ and real $\mathcal{G}_a$ multiplets of minimal supergravity \cite{Wess:1992cp,Buchbinder:1998qv}.

\subsection{$\mathcal{N}=2$ supergravity} 
\label{sec:N=2rev}

The theory preserving eight supercharges in four dimensions is denoted $\mathcal{N}=2$ supergravity. With respect to the minimal theory, the larger amount of supersymmetry constrains the interactions in a more severe way. The ungauged theory is known to arise for example as the low energy limit of the type II superstring compactified on a Calabi-Yau threefold. An orientifold of such construction may then truncate the amount of supersymmetry down to $\mathcal{N}=1$.
Below, we review the $\mathcal{N}=2$ theory following mostly the conventions of \cite{Andrianopoli:1996cm,Ceresole:1995ca}. With respect to these works, we differ by a minus sign in the definition of the Ricci scalar, and also in the signature of the spacetime metric, which we take to be mostly plus.

The main ingredients are the gravity multiplet and matter multiplets, among which we recall vector and hyper multiplets. The gravity multiplet contains the graviton, an SU$(2)_R$ doublet of gravitini $\psi_\mu^A$, with $A=1,2$, and the graviphoton. Vector multiplets contains one complex scalar, an SU$(2)_R$ doublet of fermions and one vector, $\{z,\lambda^A, A_\mu\}$. Hyper multiplets contain four real scalars and two fermions, $\{q^u, \zeta^1, \zeta^2\}$.

Scalar fields arise from two different supersymmetry representations and in fact the whole scalar manifold has a product structure
\begin{equation}
\mathcal{M} = \mathcal{S} \otimes \mathcal{Q}.
\end{equation}
The scalar fields $z^i$ of the vector multiplets, with $i=1,\dots,n_V$, are coordinates of a special K\"ahler manifold $\mathcal{S}$.\footnote{A special K\"ahler manifold \cite{Craps:1997gp} is a K\"ahler-Hodge manifold of dimension $n_V$ together with a Sp$((2n_V+2),\mathbb{R})$ vector bundle $H$ over it. The tensor bundle $H \otimes L$, where $L$ is the complex line bundle entering the K\"ahler-Hodge definition, has symplectic sections $(X^\Lambda, F_\Lambda)$ in terms of which the K\"ahler potential can be expressed as in \eqref{N=2Kaelpot}. Furthermore, the symplectic section has to be constrained as $X^\Lambda \partial_i F_\Lambda-\partial_i X^\Lambda F_\Lambda=0$. } 
The scalar fields $q^u$ of the hyper multiplets, with $i=1, \dots, 4n_H$, are coordinates of a quaternionic K\"ahler manifold $\mathcal{Q}$.\footnote{A quaternionic K\"ahler manifold is a $4n_H$-dimensional manifold together with an SU$(2)$ bundle such that its curvature is proportional to the HyperK\"ahler form. Despite the name, a quaternionic K\"ahler manifold is not K\"ahler, for its HyperK\"ahler form is not strictly closed, but only covariantly closed.}
Due to the product structure, the two manifolds are independent and do not mix. 

Interactions on $\mathcal{S}$ are fixed by specifying symplectic sections $(X^\Lambda, F_\Lambda)$, with $\Lambda=0,1,\dots, n_V$, or possibly a prepotential $F=F(X)$, such that $F_\Lambda = \partial_\Lambda F$ (this relation might not hold in all symplectic frames). The physical scalars of the vector multiplets are recovered as local coordinates on $\mathcal{S}$, namely
\begin{equation}
\label{normcoord}
z^i = \frac{X^i}{X^0}, \qquad X^0 \equiv 1.
\end{equation}
Given the symplectic section, one can construct the K\"ahler potential
\begin{equation}
\label{N=2Kaelpot}
K = - \log i \left(\bar X^\Lambda F_\Lambda - X^\Lambda \bar F_\Lambda\right).
\end{equation}
It may also be useful to introduce the sections
\begin{equation}
(L^\Lambda, M_\Lambda) = e^{\frac K2}(X^\Lambda, F_\Lambda),
\end{equation}
such that $\bar L^\Lambda M_\Lambda- L^\Lambda \bar M_\Lambda=-i$. They are covariantly-holomorphic and their covariant derivatives 
\begin{equation}
f^\Lambda_i = D_i L^\Lambda = \left(\partial_i +\frac12 \partial_iK\right)L^\Lambda, \qquad h_{i \, \Lambda} = D_i M_\Lambda = \left(\partial_i + \frac12 \partial_i K\right)M_\Lambda,
\end{equation}
can be employed as vielbeins to construct the pullback of the K\"ahler metric
\begin{equation}
\label{ULambdaSigma}
g^{i \bar \jmath} f_i^\Lambda f_{\bar \jmath}^\Sigma \equiv U^{\Lambda \Sigma} = -\frac12 ({\rm Im}\mathcal{N}^{-1})^{\Lambda \Sigma}-\bar L^\Lambda L^\Sigma.
\end{equation}
Here, we introduced the gauge kinetic function ${\rm Im}\mathcal{N}_{\Lambda \Sigma}$, which is the imaginary part of the complex symmetric matrix $\mathcal{N}_{\Lambda \Sigma}$ such that $M_\Lambda= \mathcal{N}_{\Lambda \Sigma}L^\Sigma$, $h_{i \Lambda}= \bar{\mathcal{N}}_{\Lambda \Sigma}f_i^\Sigma$ and ${\rm Im}\mathcal{N}_{\Lambda \Sigma}L^\Lambda \bar L^\Sigma = -\frac12$.

Interactions on $\mathcal{Q}$ are fixed by the quaternionic metric $h_{uv} = \mathcal{U}_u^{A\alpha}\mathcal{U}_v^{B\beta}\epsilon_{AB}C_{\alpha \beta}$, where $\epsilon_{AB}$ and $C_{\alpha\beta}$ are metrics on  SU$(2)_R$ and Sp$(2n_H)$ respectively. We are not going to use much quaternionic geometry in what follows, but we refer the interested reader to \cite{Andrianopoli:1996cm} for more details.

The gauged theory can be constructed as a deformation of the ungauged one. To preserve the same amount of supercharges, the deformation affects the supersymmetry transformations of fermions, up to order $g$, with $g$ the deformation parameter, and also the Lagrangian, up to order $g^2$. In particular, the term at order $g^2$ is the scalar potential. Concretely, the gauging is performed by introducing two sets of moment maps, one on the manifold $\mathcal{S}$ and one on $\mathcal{Q}$, respectively a singlet and a SU$(2)_R$ triplet,
\begin{equation}
\mathcal{P}^0_\Lambda(z,\bar z), \qquad\qquad \mathcal{P}^{\rm x}_{\Lambda}(q,\bar q), \qquad {x}=1,2,3,
\end{equation}
such that their derivatives are the Killing vectors
\begin{equation}
k^i_\Lambda = i g^{i \bar \jmath}\partial_{\bar \jmath}\mathcal{P}^0_\Lambda, \qquad k^u_\Lambda = \frac16 \Omega^{{x},uv}\nabla_v \mathcal{P}^x_\Lambda.
\end{equation}
Here, we have $\nabla_u \mathcal{P}^x_\Lambda = \partial_u \mathcal{P}^{x}_\Lambda + \epsilon^{xyz}\omega^y_u \mathcal{P}^z_\Lambda$, where $\omega^x$ is the connection of the SU$(2)$ bundle with curvature $\Omega^x$. The Killing vectors furnish a representation of the gauge algebra $[k_\Lambda, k_\Sigma] = - {f_{\Lambda\Sigma}}^\Gamma k_\Gamma$ (the minus sign difference with respect to the $\mathcal{N}=1$ discussion is due to conventions) and the prepotentials have to satisfy the equivariance consistency conditions
\begin{align}
&i g_{i \bar \jmath}(k^{i}_\Lambda k^{\bar \jmath}_{\Sigma}-k^{i}_\Sigma k^{\bar \jmath}_{\Lambda}) =-{f_{\Lambda\Sigma}}^\Gamma \mathcal{P}^0_{\Gamma},\\
\label{equivhyper}
&2k_\Lambda^u k_\Sigma^v \Omega^x_{uv} = -{f_{\Lambda\Sigma}}^\Gamma \mathcal{P}^x_\Gamma-\epsilon^{xyz}\mathcal{P}^y_\Lambda \mathcal{P}^z_{\Sigma}.
\end{align}
 
The bosonic sector of the theory reads (we set the deformation parameter $g=1$ in what follows)
\begin{equation}
\begin{aligned}
e^{-1}\mathcal{L} = &\frac12 R - g_{i \bar \jmath}D_\mu z^i D^\mu \bar z^{\bar \jmath} - h_{uv}D_\mu q^u D^\mu q^v\\
& + {\rm Im} \mathcal{N}_{\Lambda \Sigma}F_{\mu\nu}^{\Lambda}F^{\mu\nu \, \Sigma} + \frac12 {\rm Re}\mathcal{N}_{\Lambda \Sigma}\epsilon^{\mu\nu\rho\sigma} F_{\mu\nu}^{\Lambda} F_{\rho\sigma}^{\Sigma}-V,
\end{aligned}
\end{equation}
where the scalar potential is
\begin{equation}
\label{VN=2full}
V = \left(g_{i\bar\jmath}k^i_\Lambda k^{\bar\jmath}_\Sigma+ 4h_{uv}k^u_\Lambda k^v_\Sigma\right)\bar L^{\Lambda}L^\Sigma  + \left(U^{\Lambda \Sigma}-3 \bar L^{\Lambda}L^\Sigma\right)\mathcal{P}^x_\Lambda \mathcal{P}^x_{\Sigma}.
\end{equation}
Notice that the matrix ${\rm Im}\mathcal{N}_{\Lambda \Sigma}$ is negative definite, as it can be deduced from the kinetic term of the vector fields.

The only piece of information from the fermionic sector that we are going to employ extensively is the gravitino covariant derivative, 
\begin{equation}
\begin{aligned}
\label{DcovgravN2}
D_\mu \psi_\nu^A &=\partial_\mu \psi_\nu^A -\frac14 \omega_\mu^{ab}\gamma_{ab}\psi_\nu^A -\frac i2 \left(v_\mu +A^\Lambda_\mu \mathcal{P}^0_\Lambda\right)\psi_\nu^A + \left({\omega_\mu^A}_B +\frac i2 A^\Lambda_\mu \mathcal{P}^x_\Lambda {(\sigma^x)^A}_B\right) \psi^B_\nu\\
&=\dots-\frac i2 A^\Lambda_\mu \mathcal{P}^0_\Lambda \psi_\nu^A+\frac i2  A^\Lambda_\mu {\mathcal{P}^x}_\Lambda {(\sigma^x)^A}_B \psi^B_\nu,
\end{aligned}
\end{equation}
where in the second line we highlighted the terms which are most relevant for our analysis. Here, $v={\rm Im}(\partial K)$ is the connection on the line bundle associated to a K\"ahler-Hodge manifold, while ${\omega^A}_B = \frac i2 \omega^x {(\sigma^x)^A}_B$ is the connection on the SU$(2)$ bundle characterizing the quaternionic K\"ahler manifold.

For completeness, we report also the supersymmetry transformations of the fermions
\begin{align}
\delta \psi_{A\mu}&= D_\mu \epsilon_A - i S_{AB}\gamma_\mu \epsilon^B + \epsilon_{AB} T_{\mu\nu}^{-}\gamma^\nu \epsilon^B + \mathcal{O}(\psi,\lambda,\zeta),\\
\delta \lambda^{iA} &= W^{iAB}\epsilon_B +i \gamma^\mu D_\mu z^i \epsilon^A -G_{\mu\nu}^{i-}\gamma^{\mu\nu}\epsilon^A + \mathcal{O}(\psi,\lambda,\zeta),\\
\delta \zeta_\alpha &= N_\alpha^A\epsilon_A + i\mathcal{U}_{u}^{B\beta}\gamma^\mu D_\mu q^u \epsilon^A \epsilon_{AB}C_{\alpha \beta}+ \mathcal{O}(\psi,\lambda,\zeta),
\end{align}
where we can distinguish the so called fermionic shifts
\begin{align}
\label{gravmassmatrix}
S_{AB} &=\frac i2 {(\sigma^x)_A}^C\epsilon_{BC} \mathcal{P}^x_\Lambda L^\Lambda,\\
W^{iAB} &=\epsilon^{AB}k_\Lambda^i \bar L^\Lambda+i{(\sigma^x)_C}^B\epsilon^{CA}\mathcal{P}^x_\Lambda g^{i\bar\jmath}\bar f^\Lambda_{\bar \jmath},\\
N^A_\alpha &=2 \mathcal{U}^A_{\alpha u}k^u_\Lambda \bar L^\Lambda.
\end{align}
Here, the quantity $S_{AB}$ is the gravitino mass matrix. Furthermore, we introduced the objects
\begin{align}
\label{Tpm}
T_{\mu\nu}^- &= 2i{\rm Im}\mathcal{N}_{\Lambda \Sigma} L^\Lambda F^{\Sigma -}_{\mu\nu} + \mathcal{O}(\psi,\lambda,\zeta),\\
G_{\mu\nu}^{i-} &= - g^{i\bar \jmath} {\rm Im}\mathcal{N}_{\Lambda \Sigma} \bar f^\Lambda_{\bar\jmath}F_{\mu\nu}^{\Sigma -}+ \mathcal{O}(\psi,\lambda,\zeta),
\end{align}
while the analogous quantities $T_{\mu\nu}^- $ and $G_{\mu\nu}^{i-}$ can be found by complex conjugation.

Notice that the supersymmetry transformations link the gravity multiplet, namely the gravitino, to a certain combination $T_{\mu\nu}^{\pm}$ of vector field strengths and special K\"ahler scalar fields. The orthogonal combination $G^{i \pm }_{\mu\nu}$ is instead entering the supersymmetry transformations of the gaugini $\lambda^{iA}$. Orthogonality follows from the special geometry identity ${\rm Im}\mathcal{N}_{\Lambda \Sigma}L^\Lambda f^\Sigma_i=0$. Due to this structure, even if the vector $A^0_\mu$ is usually referred to as graviphoton, when supergravity is coupled to matter it might be more convenient to really think of the graviphoton as the linear combination of vector and scalar fields giving the field strengths entering the supersymmetry transformations of the gravitino in the gravity multiplet. Accordingly, the matter vector fields will be the orthogonal combination, entering the supersymmetry transformations of the matter multiplets.

\section{WGC and supergravity EFTs}
\label{sec:WGCvs}

The Weak Gravity Conjecture (WGC) \cite{Arkani-Hamed:2006emk} is one of the first swampland conjectures to be proposed and it is currently one of the most established. There are various ways of formulating it, but perhaps the simplest is the statement that gravity has to be the weakest force. According to the so called electric version of the WGC, any effective theory coupled to gravity and with an abelian gauge group should contain a state whose mass is smaller than the product of the gauge coupling and the charge in Planck units, times a numerical coefficient which can be calculated. In the following, we will employ the so called magnetic formulation of the WGC, which states that the ultraviolet cutoff $\Lambda_{UV}$ of an effective theory with an abelian gauge group and coupled to gravity is bounded from above by the gauge coupling $g$ in Planck units, namely (in four dimensions)
\begin{equation}
\label{magneticWGC}
\Lambda_{UV} \lesssim g M_P.
\end{equation}

The magnetic WGC is the prototype example of swampland conjecture in the following sense. Naively, one would assume that the ultraviolet cutoff of an effective theory coupled to gravity is the Planck scale, namely the scale at which the gravitational interaction becomes strongly coupled.\footnote{In presence of a high number of light species, this statement might not be correct and one should rather use the species scale as cutoff \cite{Dvali:2007hz,Dvali:2007wp,Dvali:2009ks,Dvali:2010vm,Dvali:2012uq}.} However, what the magnetic WGC is telling us is that this naive expectation is indeed not correct. The cutoff of the effective theory is lower than expected, due to quantum gravity effects. One can think of \eqref{magneticWGC} as a constraint on a given effective theory that is not obvious from a bottom-up perspective, but that is required for having a consistent coupling to gravity in the ultraviolet regime. In other words, one cannot really probe the theory up to the Planck scale, but rather up to the scale dictated by the gauge coupling in Planck units.\footnote{The gauge coupling is a running coupling and thus its value depends on the energy scale at which it is measured. As for the magnetic WGC statement, in \cite{Arkani-Hamed:2006emk} it is suggested that $g$ appearing in \eqref{magneticWGC} has to be measured at the scale $\Lambda_{UV}$, which is a defining scale of the effective theory.}

The WGC has been originally formulated in flat space and, as such, it might not directly apply to a curved background. Therefore, care is needed when trying to use the WGC in, say, anti-de Sitter or de Sitter space. To proceed, we will assume that some formulation of the WGC on a curved background exists, for example the one proposed by \cite{Huang:2006hc} (see also \cite{Antoniadis:2020xso}), and it is such that the deviation from the flat space formulation is suppressed by the Hubble or (anti-)de Sitter radius. In the regime of validity of the supergravity approximation, this radius has to be large and thus the expected corrections should be negligible. This will be the main working assumption for the results reviewed in the present section.

In the following, we are going to show that in large classes of supergravity models with at least eight supercharges at the Lagrangian level, the cosmological constant is of the form (omitting numerical factors)
\begin{equation}
\label{V>g2}
|V| \geq g_{3/2}^2 M_P^2,
\end{equation}
where $g_{3/2}$ is the gravitino gauge coupling associated to the gauging of the R-symmetry. The bound is actually saturated for $\mathcal{N}=2$ anti-de Sitter vacua, while for de Sitter vacua we will need an additional assumption on the gravitino mass to arrive precisely at \eqref{V>g2}. The above relation will be then the starting point to apply swampland conjectures.

\subsection{WGC versus scale separation}

In this section, we focus on $\mathcal{N}=2$ anti-de Sitter vacua and we review the general argument given in \cite{Cribiori:2022trc} which forbids the existence of scale separation as a consequence of the WGC. 

Scale separation is a property of certain models with extra compact spacetime dimensions. It amounts to ask that the typical length scale of the non-compact dimensions, e.g.~the anti-de Sitter radius $L_{AdS}$ in our case, is (parametrically) larger than the Kaluza-Klein scale $L_{KK}$ associated to the extra dimensions. In practice, for scale separation one would like to have 
\begin{equation}
\frac{L_{KK}}{L_{AdS}} \ll 1.
\end{equation}
As such, this is a phenomenologically motivated requirement, for no experimental evidence of more than four spacetime dimensions is confirmed at present, while on the other hand critical string theory is defined in ten spacetime dimensions. 
One way to bring string theory closer to the real world is thus to assume that the extra dimensions are indeed compact and small compared to the observed ones. In particular, they are supposed to be so small that they have not been observed yet.

The vast majority of known supersymmetric anti-de Sitter backgrounds in string theory do not admit scale separation. The prototype example is perhaps the celebrated $AdS_5 \times S^5$ solution of type IIB string theory, for which it is known that the sizes of the two spaces are identified by supersymmetry. In fact, no anti-de Sitter vacuum with at least eight supercharges and with scale separation has been found in the string landscape, to the best of our knowledge; see e.g.~\cite{Lust:2004ig,Tsimpis:2012tu,Gautason:2015tig,Gautason:2018gln,Lust:2020npd,DeLuca:2021mcj,Collins:2022nux,Andriot:2022yyj} in support to this statement. This remarkable amount of evidence led to the proposal of swampland conjectures forbidding scale separation in anti-de Sitter \cite{Lust:2019zwm}, see also \cite{Blumenhagen:2019vgj,Buratti:2020kda,Emelin:2020buq,Shiu:2022oti} for interesting related works. To summarise, one could imagine a relation of the following type 
\begin{equation}
\label{LAdSLKK}
L_{AdS} \sim (L_{KK})^\alpha,
\end{equation}
for some order one parameter $\alpha$. The strong version of the conjecture in \cite{Lust:2019zwm} postulates that $\alpha=1$ and thus no scale separation is possible in anti-de Sitter.

However, certain counterexamples to $L_{AdS} \sim L_{KK}$ are known. These are a class of four-dimensional $\mathcal{N}=1$ anti-de Sitter vacua from massive IIA compactifications, usually denoted as DGKT vacua \cite{DeWolfe:2005uu} (see also \cite{Behrndt:2004km,Derendinger:2004jn}), their double T-dual versions in massless IIA/M-theory \cite{Cribiori:2021djm} and three-dimensional vacua of \cite{Farakos:2020phe,VanHemelryck:2022ynr,Farakos:2023nms} from massive IIA on $G_2$ orientifolds. Even if these models recently passed non-trivial tests \cite{Junghans:2020acz,Marchesano:2020qvg,Cribiori:2021djm,Emelin:2022cac}, their precise embedding in string theory is yet not completely understood.
In the following, we will show how a relation of the type $L_{AdS}\sim L_{KK}$ can be derived from the magnetic WGC in vacua with at least eight supercharges.

The conditions to find supersymmetric anti-de Sitter vacua in $\mathcal{N}=2$ supergravity have been studied in \cite{Hristov:2009uj} and, in general, they can be found by setting to zero the supersymmetry variations of the fermions and assuming maximal symmetry of the background. In our case, they are
\begin{equation}
k^i_\Lambda = k^u_\Lambda =\mathcal{P}^x_\Lambda f^\Lambda_i=0.
\end{equation}
The vacuum energy thus reduces to the (trace of the) gravitino mass
\begin{equation}
V_{AdS} = - 3 \bar L^\Lambda L^\Sigma \mathcal{P}^x_\Lambda \mathcal{P}^x_\Sigma.
\end{equation}
For the purposes of our analysis, we have to rewrite it in terms of the gauge couplings. While with only four supercharges such a step cannot in general be performed, thanks to the extended amount of supersymmetry it can here be performed in a model-independent manner. Taking the square of the vacuum condition $\mathcal{P}^x_\Lambda f^\Lambda_i=0$ and using the relation \eqref{ULambdaSigma}, we have
\begin{equation}
\bar L^\Lambda L^\Sigma \mathcal{P}_\Lambda^x \mathcal{P}_\Sigma^x = -\frac12 ({\rm Im}\mathcal{N}^{-1})^{\Lambda \Sigma}\mathcal{P}^x_\Lambda \mathcal{P}^x_\Sigma
\end{equation}
and thus
\begin{equation}
V_{AdS} = \frac32 ({\rm Im}\mathcal{N}^{-1})^{\Lambda \Sigma}\mathcal{P}^x_\Lambda \mathcal{P}^x_\Sigma.
\end{equation}
Given that in special geometry one has $({\rm Im}\mathcal{N}^{-1})^{\Lambda \Sigma}\mathcal{P}_{\Lambda}^0\mathcal{P}_\Sigma^0 = -2g_{i \bar \jmath}k_\Lambda^i k_\Sigma^{\bar \jmath} \bar L^\Lambda L^\Sigma$, which can be derived from $\mathcal{P}^0_\Lambda L^\Lambda=0$, and that this vanishes on the vacuum, we can freely supplement the expression for $V_{AdS}$ with an additional term to get
\begin{equation}
\label{VAdS=QQ}
V_{AdS} = \frac32 ({\rm Im}\mathcal{N}^{-1})^{\Lambda \Sigma}\left(\mathcal{P}^0_\Lambda\mathcal{P}^0_\Sigma+\mathcal{P}^x_\Lambda \mathcal{P}^x_\Sigma\right) \equiv 3 ({\rm Im}\mathcal{N}^{-1})^{\Lambda \Sigma} \,{\rm Tr}\, \mathcal{Q}_\Lambda \mathcal{Q}_\Sigma,
\end{equation}
where we defined the SU$(2)$ charge matrix
\begin{equation}
2{\mathcal{Q}_{\Lambda\,A}}^B = \mathcal{P}^0_\Lambda \delta_A^B + \mathcal{P}^x_\Lambda {(\sigma^x)_A}^B,
\end{equation}
such that
\begin{equation}
{\rm Tr}\, \mathcal{Q}_\Lambda \mathcal{Q}_\Sigma = \frac12 \left(\mathcal{P}^0_\Lambda\mathcal{P}^0_\Sigma+\mathcal{P}^x_\Lambda \mathcal{P}^x_\Sigma\right).
\end{equation}

The expression \eqref{VAdS=QQ} is on the right track towards \eqref{V>g2}, since it expresses the cosmological constant as a product of the gauge kinetic function and of the charge, without any other quantity entering the relation. However, the WGC is formulated in terms of an abelian gauge coupling associated to a canonically normalized field. Thus, to make contact with the WGC, we should identify an unbroken abelian gauge group on the vacuum and canonically normalize the gauge field associated to it.

To identify the unbroken group on the vacuum we take advantage of the results of \cite{Lust:2017aqj}. According to that work, on a supersymmetric anti-de Sitter vacuum the unbroken gauge group branches into two factors, one of which, $H^g_R$, is gauged by the graviphotons of the theory. This statement is fairly independent from the number of dimensions and of preserved supercharges. In four dimensions, one has \cite{Lust:2017aqj}
\begin{equation}
H^g_R = {\rm SO}(\mathcal{N}),
\end{equation}
implying that $\mathcal{N}=2$ vacua have $H^g_R = {\rm U}(1)_R$. Thus, in the vacua under investigation there is always an unbroken abelian gauge group of the full gauge group and we can proceed by applying the WGC with respect to it.

By definition, $H_R^g$ is gauged by the graviphoton(s). In this context, what is meant by graviphoton is not really $A^0_\mu$, but rather the linear combination of vector fields (and scalars) whose field strengths enter the supersymmetry transformation of the gravitini, as commented at the of section \ref{sec:N=2rev}. Thus, the candidate vector field gauging $H_R^g$ is in fact a linear combinations of vectors and scalars. Without loss of generality, we introduce a general linear combination of vector fields (we suppress the spacetime index when obvious)
\begin{equation}
\tilde A = \Theta_\Lambda A^\Lambda.
\end{equation}
The vector $\tilde A$ will be the one associated to the WGC.
Here, $\Theta_\Lambda$ are arbitrary coefficients; one can think of them more or less as the vacuum expectation values of the scalar fields constructing $T_{\mu\nu}^\pm$ in \eqref{Tpm}. In fact, in concrete models the coefficients $\Theta_\Lambda$ can be identified by asking that $\tilde A$ gauges $H_R^g$. This fixes $\Theta_\Lambda$ up to an overall normalization which can then be chosen my matching with the vacuum energy.

Next, we split the whole set of vector fields, charges and moment maps into objects which are parallel and orthogonal with respect to the direction singled out by $\tilde A$. This can be done with the projectors used in \cite{Cribiori:2020use,Cribiori:2022trc}. We have (indices $\Lambda, \Sigma, \dots$ are raised and lowered with ${\rm Im}\mathcal{N}_{\Lambda \Sigma}$)
\begin{equation}
\begin{aligned}
A_\Lambda &= A^\parallel_\Lambda + A^\perp_\Lambda  
= \frac{\Theta_\Lambda}{\Theta^2} \tilde A + A^\perp_\Lambda,
\end{aligned}
\end{equation}
where we denoted $\Theta^2 = \Theta_\Lambda ({\rm Im}\mathcal{N}^{-1})^{\Lambda \Sigma} \Theta_\Sigma$.
Similarly, we split the combination $A^\Lambda \mathcal{Q}_\Lambda$ entering the gravitino covariant derivative (we omit SU$(2)$ indices)
\begin{equation}
A_\Lambda \mathcal{Q}^\Lambda = A_\Lambda \mathcal{Q}^{\parallel \Lambda} +  A_\Lambda \mathcal{Q}^{\perp \Lambda} = \tilde{\mathcal{Q}} \tilde A +  A_\Lambda \mathcal{Q}^{\perp \Lambda},
\end{equation}
where in the last step we defined the gravitino charge
\begin{equation}
{\tilde{\mathcal{Q}}_A}^{\,\,\,\,B}= \frac{({\rm Im}\mathcal{N}^{-1})^{\Lambda \Sigma}\Theta_\Sigma}{\Theta^2} {\mathcal{Q}^\parallel_{\Lambda\, A}}^B, \qquad {\mathcal{Q}^\parallel_{\Lambda\, A}}^B = \Theta_\Lambda  {\tilde{\mathcal{Q}}_A}^{\,\,\,\,B}.
\end{equation}
We split also the vector fields kinetic terms 
\begin{equation}
\begin{aligned}
{\rm Im}\mathcal{N}_{\Lambda \Sigma} F^\Lambda(A) \wedge *  F^\Sigma(A) &= {\rm Im} \mathcal{N}_{\Lambda \Sigma}\frac{\Theta^\Lambda}{\Theta^2}\frac{ \Theta^\Sigma}{\Theta^2} F (\tilde A) \wedge * F(\tilde A) + {\rm Im}\mathcal{N}_{\Lambda \Sigma} F^\Lambda(A^\perp) \wedge *  F^\Sigma(A^\perp) \\
&=\frac{1}{\Theta^2} F (\tilde A) \wedge * F(\tilde A) + {\rm Im}\mathcal{N}_{\Lambda \Sigma} F^\Lambda(A^\perp) \wedge *  F^\Sigma(A^\perp),
\end{aligned}
\end{equation}
from which we identify the (abelian) gravitino gauge coupling
\begin{equation}
g_{3/2}^2 =- \frac14 \Theta^2 =-\frac14 \Theta_\Lambda ({\rm Im}\mathcal{N}^{-1})^{\Lambda \Sigma} \Theta_\Sigma.
\end{equation}
Recall that the matrix ${\rm Im}\mathcal{N}_{\Lambda \Sigma}$ is negative definite.

Taking all of these redefinitions into account, we can eventually recast the scalar potential as
\begin{equation}
\begin{aligned}
V_{AdS} &= 3 ({\rm Im}\mathcal{N}^{-1})^{\Lambda \Sigma} \,{\rm Tr}\, \mathcal{Q}_\Lambda \mathcal{Q}_\Sigma\\
&=3({\rm Im}\mathcal{N}^{-1})^{\Lambda \Sigma} \,{\rm Tr}\, \left(\mathcal{Q}_\Lambda^\parallel \mathcal{Q}_\Sigma^\parallel+\mathcal{Q}_\Lambda^\perp \mathcal{Q}_\Sigma^\perp\right)\\
&\leq 3 ({\rm Im}\mathcal{N}^{-1})^{\Lambda \Sigma} \,{\rm Tr}\, \mathcal{Q}_\Lambda^\parallel \mathcal{Q}_\Sigma^\parallel\\
&=-12 \,g_{3/2}^2\, {\rm Tr}\, \tilde{\mathcal{Q}} \tilde{\mathcal{Q}},
\end{aligned}
\end{equation}
or in absolute value
\begin{equation}
\label{VAdSg2}
|V_{AdS}| \geq 12 g_{3/2}^2 {\rm Tr}\, \tilde{\mathcal{Q}} \tilde{\mathcal{Q}}.
\end{equation}
Eventually, we derived \eqref{V>g2} as desired.

Actually, for $\mathcal{N}=2$ vacua the bound is saturated. Indeed, the contributions along $A^\perp$ would be associated to vector fields gauging an orthogonal direction with respect to $U(1)_R$, i.e~the abelian factor with respect to which we apply the WGC, but still inside the group $H_R^g$. Since in these vacua $H_R^g=U(1)$, there cannot be any such orthogonal contribution. On vacua with more preserved supercharges, the group $H_R^g$ is enlarged and thus in general one finds an inequality as above.

The relation \eqref{VAdSg2} is the main formula from which we can draw our conclusions on scale separation. In what follows, we assume that some form of charge quantization is implemented, in the sense that $ {\rm Tr}\, \tilde{\mathcal{Q}} \tilde{\mathcal{Q}}$ is quantized and cannot be arbitrarily small.\footnote{This assumption is not needed if one restores the charge in the original formula of the WGC, namely if one uses $\Lambda_{UV}\lesssim gq M_P$.} Therefore, the only parameter which can in principle vary arbitrarily is the gauge coupling $g_{3/2}$.
A first observation is that, in the limit $g_{3/2} \to 0$, assuming this is smooth, we recover a global (R-)symmetry. Since there cannot be global symmetries in quantum gravity, such a limit should not be allowed and the cosmological constant cannot be arbitrarily small. This has been recently pointed out in \cite{Montero:2022ghl}. However, we can argue that the relation \eqref{VAdSg2} is problematic already as it is, without the need to take any limit. Indeed, if we enforce the magnetic WGC \eqref{magneticWGC}, we have
\begin{equation}
|V_{AdS}| \gtrsim \Lambda_{UV}^2 M_P^2,
\end{equation}
which is telling us that the cosmological constant is quantized in terms of the ultraviolet cutoff of the theory. In particular, the two scales cannot be decoupled since there is no additional parameter which can be used to disentangle them. If the ultraviolet cutoff dictated by the WGC is the Kaluza-Klein scale $L_{KK}$, as it is natural to assume in supergravity, then the relation above is in fact
\begin{equation}
\frac{L_{KK}}{L_{AdS}} \gtrsim  1,
\end{equation}
which forbids scale separation in these vacua. Thus, supersymmetric anti-de Sitter vacua with (at least) eight supercharges are not genuinely four-dimensional effective theories if the WGC holds. Rather, they should be understood as higher dimensional theories. We refer to \cite{Cribiori:2022trc} for a discussion on how to extend the result to more than eight preserved supercharges and for an explicit proof in the maximal theory in four dimensions. The extension to five dimensions has been recently worked out in \cite{Cribiori:2023ihv}.

\subsection{WGC versus de Sitter}

A slightly modified version of the argument above against scale separation in anti-de Sitter can be formulated and used to constrain effective theories with de Sitter critical points and with a vanishing gravitino mass on the vacuum. This argument appeared originally in \cite{DallAgata:2021nnr,Emelin:2022wft,Cribiori:2022sxf} and is reviewed in the present section together with some examples.

The motivation to constrain the viability of de Sitter vacua, and more in general critical points, in effective supergravity theories is phenomenological, for one of the possible explanations for the positive value of the dark energy density measured today is a cosmological constant. On the other hand, constructing de Sitter vacua in string theory seems extremely challenging, at the very least. Even the most studied scenarios, namely KKLT and LVS \cite{Kachru:2003aw,Balasubramanian:2005zx}, are not free from criticisms, see e.g.~\cite{Danielsson:2018ztv,Gautason:2018gln,Gao:2020xqh,Junghans:2022exo,Gao:2022fdi,Junghans:2022kxg,Blumenhagen:2022dbo} for recent works. The absence of fully understood examples has once more led to the formulation of swampland conjectures forbidding de Sitter vacua in quantum gravity, such as those of \cite{Obied:2018sgi,Garg:2018reu,Andriot:2018wzk,Ooguri:2018wrx,Bedroya:2019snp}. 
It is worth to observe however, that, imposing restrictions that have been deduced by the properties of well-controlled regions of moduli space to hold also in the interior may be misleading; 
such an issue has indeed been raised in \cite{Cicoli:2021fsd,Cicoli:2021skd} for the study of late-time cosmology. 
The de Sitter criterion has been studied also directly in four-dimensional $\mathcal{N}=1$ supergravity in \cite{Ferrara:2019tmu}. 
In the following, we show that, similarly to scale separation, the WGC can be used to rule out certain de Sitter critical points in gauged supergravity without the need to enforce other conjectures.

Looking at the complete scalar potential \eqref{VN=2full} and assuming the gravitino mass to be vanishing on the background, we have a manifestly positive expression of the form
\begin{equation}
\begin{aligned}
\label{VdSWGC1}
V_{dS} &= -\frac12 \left({\rm Im}\mathcal{N}^{-1}\right)^{\Lambda \Sigma} \left(\mathcal{P}^0_\Lambda\mathcal{P}^0_\Sigma+\mathcal{P}^x_\Lambda \mathcal{P}^x_\Sigma\right) + 4 h_{uv}k^u_\Lambda k^v_\Sigma \bar L^\Lambda L^\Sigma\\
&\geq  -\frac12 \left({\rm Im}\mathcal{N}^{-1}\right)^{\Lambda \Sigma} \left(\mathcal{P}^0_\Lambda\mathcal{P}^0_\Sigma+\mathcal{P}^x_\Lambda \mathcal{P}^x_\Sigma\right)\\
&=-\left({\rm Im}\mathcal{N}^{-1}\right)^{\Lambda \Sigma} {\rm Tr}\, \mathcal{Q}_\Lambda \mathcal{Q}_\Sigma,
\end{aligned}
\end{equation}
where we used \eqref{ULambdaSigma} and also $g_{i \bar\jmath}k^i_\Lambda k^{\bar \jmath}_\Sigma \bar L^\Lambda L^\Sigma=-\frac12 ({\rm Im}\mathcal{N}^{-1})^{\Lambda \Sigma}\mathcal{P}^0_\Lambda \mathcal{P}^0_\Sigma$, which can be derived from $\mathcal{P}^0_\Lambda L^\Lambda=0$. This form of the scalar potential is analogous to \eqref{VAdS=QQ}.

 We can now repeat a similar analysis as for the anti-de Sitter case. One important difference concerns the presence of an unbroken abelian group at the critical point, which is not guaranteed in de Sitter. In fact, the unbroken gauge group can in general be different or even absent. 
 Besides, a subgroup of the gauge group must be in any case broken, otherwise supersymmetry would be completely preserved. To proceed, we need to assume that part of the gauge group is preserved at the critical point. In case this part contains an abelian factor, we can apply the WGC directly to it. In case, this does not happen, we can still apply the WGC with respect to any of the Cartan generators of the unbroken gauge algebra.

Keeping this in mind, we can proceed as for anti-de Sitter, namely by splitting all quantities with a symplectic index $\Lambda=0,1,\dots,n_V$ in directions parallel and orthogonal with respect to the WGC one. Using the same redefinitions as before for vectors, charges, and gauge coupling, we can then write \eqref{VdSWGC1} as
\begin{equation}
\begin{aligned}
V_{dS} &\geq -\left({\rm Im}\mathcal{N}^{-1}\right)^{\Lambda \Sigma} {\rm Tr}\, \mathcal{Q}_\Lambda \mathcal{Q}_\Sigma\\
&=-({\rm Im}\mathcal{N}^{-1})^{\Lambda \Sigma} \,{\rm Tr}\, \left(\mathcal{Q}_\Lambda^\parallel \mathcal{Q}_\Sigma^\parallel+\mathcal{Q}_\Lambda^\perp \mathcal{Q}_\Sigma^\perp\right)\\
&\geq - ({\rm Im}\mathcal{N}^{-1})^{\Lambda \Sigma} \,{\rm Tr}\, \mathcal{Q}_\Lambda^\parallel \mathcal{Q}_\Sigma^\parallel\\
&=4 \,g_{3/2}^2\, {\rm Tr}\, \tilde{\mathcal{Q}} \tilde{\mathcal{Q}}.
\end{aligned}
\end{equation}
Eventually, we derived \eqref{V>g2} for de Sitter critical points with a vanishing gravitino mass. The argument extends directly to gravitini with parametrically small mass compared to the Hubble scale.

This formula is again the main relation from which we can draw our conclusions. 
First, one can see easily that a parametrically small positive cosmological constant would lead to the restoration of a global symmetry as $g_{3/2}^2 \to 0$ (assuming smoothness of the limit). This is believed not to be possible in quantum gravity. Second, one can see that the relation
\begin{equation}
\label{VdS>=g2}
V_{dS}\geq 4 \,g_{3/2}^2\, {\rm Tr}\, \tilde{\mathcal{Q}} \tilde{\mathcal{Q}}
\end{equation}
is problematic as it is, if the WGC holds. To this purpose, recall that in de Sitter the Hubble scale, ${H} \simeq \sqrt{V_{dS}}$, is a natural proxy for the infrared cutoff, ${H} \simeq \Lambda_{IR}$, for it indicates the longest length that can be measured. Applying the WGC to \eqref{VdS>=g2}, we thus have
\begin{equation}
\label{LIR=LUV}
\Lambda_{IR}^2 \simeq V_{dS} \gtrsim \Lambda_{UV}^2.
\end{equation}
This is worrisome for an effective theory, for it means that the latter is not protected against corrections suppressed by $\Lambda_{UV}$. Indeed, a favorable situation in which any effective theory should be defined is $\Lambda_{IR} \ll \Lambda_{UV}$, but this is clearly in contradiction with \eqref{LIR=LUV}. We conclude that these de Sitter critical points of gauged supergravity with vanishing gravitino mass cannot be considered as consistent low energy effective theories and rather belong to the swampland.

\subsubsection{Examples}

The condition of vanishing gravitino mass on a de Sitter critical point might seem exotic, but it is really not. Indeed, it is met in several explicit models, including all known stable de Sitter vacua in $\mathcal{N}=2$ supergravity. In this sense, the results here reviewed suggest that the most promising chance to get de Sitter vacua in four-dimension is in models with at most four-supercharges. This is compatible with the top-down analysis of \cite{Andriot:2021rdy,Andriot:2022way,Andriot:2022yyj}. Before looking at stable vacua, let us review what is probably the simplest example in which the argument above can be run.

\paragraph{Pure Fayet--Iliopoulos terms.}
The relation between the magnetic WGC and de Sitter critical points has been pointed out first in \cite{Cribiori:2020wch}, for a very simple model in $\mathcal{N}=1$ supergravity. This is the so called Freedman model, namely a pure Fayet--Iliopoulos term. In the language of section \ref{sec:N=1rev}, this is a model without chiral multiplets, but with just one vector multiplet and a constant moment map $\mathcal{P}_\Lambda = \xi \delta^1_\Lambda$, leading to a scalar potential
\begin{equation}
\label{VFIN=1}
V_{FI} = \frac12 g^2 \xi^2.
\end{equation}
Since there is no superpotential, the gravitino mass is identically vanishing. Indeed, this is required by consistency of the model itself, for a non-vanishing superpotential would break explicitly the U$(1)_R$ symmetry which is here gauged.\footnote{Fayet--Iliopoulos terms associated to the gauging of a non-R-symmetry have been constructed in \cite{Cribiori:2017laj} and subsequently developed e.g.~in \cite{Kuzenko:2018jlz,Antoniadis:2018cpq,Antoniadis:2018oeh,Antoniadis:2019hbu}. As such, they allow in principle for a non-vanishing gravitino mass and thus might evade the argument ruling out de Sitter critical points as a consequence of the WGC.} Therefore, the argument presented above applies and the simple Freedman model without chiral multiplets is in the swampland.

Notice that one can also arrive at a related conclusion by asking that there be no global symmetries in quantum gravity. In the scalar potential \eqref{VFIN=1} $\xi$ plays the role of a charge and as such it should be quantized \cite{Seiberg:2010qd,Distler:2010zg,Hellerman:2010fv}. Then, the only way to get a parametrically small cosmological constant is by tuning the gauge coupling $g$ to be small as well, but this would lead to the restoration of a global symmetry. Indeed, a connection between pure Fayet--Iliopoulos terms and the presence of a global symmetry has been pointed out already in \cite{Komargodski:2009pc}. Given that swampland conjectures are believed to be related to one another, one may expect that de Sitter critical points can be constrained also by the WGC. In fact, one can understand the WGC as an obstruction in restoring a global symmetry.

A simple model with pure Fayet--Iliopoulos terms in $\mathcal{N}=2$ supergravity can be analysed in a similar way and it actually fits our discussion both for de Sitter critical points and for anti-de Sitter vacua with(out) scale separation. In fact, the de Sitter critical point to be discussed is unstable, but the lesson one learns is valid and it can be applied to stable de Sitter vacua as well. The model appears as Exercise 6.1 in the textbook \cite{Lauria:2020rhc}, where the reader can find more details. Here, we simply recall the few ingredients that we need for our discussion in relation to the WGC.

We consider $\mathcal{N}=2$ supergravity coupled to a single vector multiplet with interactions governed by a prepotential $F(X)=-\frac i4 \left((X^0)^2-(X^1)^2\right)$. The complex scalar spans a special K\"ahler manifold of the type
\begin{equation}
\mathcal{S} = \frac{{\rm SU}(1,1)}{{\rm U}(1)}
\end{equation}
and its associated K\"ahler potential is $K=-\log \left(1-z \bar z\right)$. We can calculate the matrix $U^{\Lambda \Sigma}$ in \eqref{ULambdaSigma} and the (inverse) gauge kinetic matrix 
\begin{equation}
\label{Umatrixexample}
U^{\Lambda \Sigma} = \frac{1}{1-z \bar z} \left(\begin{array}{cc} z\bar z & \bar z\\ z & 1\end{array}\right), \qquad \left({\rm Im}\mathcal{N}^{-1}\right)^{\Lambda \Sigma} = - \frac{2}{1 - z\bar z} 
\begin{pmatrix}
1 + z \bar z & z + \overline z \\ z + \overline z & 1 + z\bar z
\end{pmatrix} \,,
\end{equation}
while the sections are $L^\Lambda = \frac{1}{\sqrt{1-z \bar z}}\left(1,z\right)^T$. Notice that at $z=0$, which is going to be the critical point in the models considered below, the gauge kinetic matrix becomes diagonal, ${\rm Im}\mathcal{N}_{\Lambda \Sigma}|_{z=0} = -\frac12 {\rm diag}(1,1)$.
The pure Fayet-Iliopoulos term arises from a gauging with constant moment maps $\mathcal{P}^x_\Lambda$. To satisfy the equivariance condition \eqref{equivhyper}, they need to point towards the same SU$(2)_R$ direction. Concretely, we choose the non-vanishing moment maps
\begin{equation}
\label{PFIexample}
\mathcal{P}^x_0 = \xi_0 \,\delta^{3x},\qquad \mathcal{P}^x_1 = \xi_1\, \delta^{3x}.
\end{equation}
Since the killing vectors are vanishing, the only term contributing to the scalar potential \eqref{VN=2full} is the last. Using \eqref{Umatrixexample} and \eqref{PFIexample}, it can be calculated to be
\begin{equation}
V = \frac{1}{1-z\bar z} \left[\xi_0^2 (z\bar z-3) -2 \xi_0 \xi_1 (z + \bar z) + \xi_1^2 (1- 3z\bar z)\right].
\end{equation}
Recall that from the gravitino covariant derivative \eqref{DcovgravN2} one can understand $\xi_0$ and $\xi_1$ in terms of charges and gauge couplings of the gravitino under the vectors $A^0_\mu$ and $A^1_\mu$ respectively. Depending on their values, there are two illustrative cases to be distinguished.
\begin{itemize} 

\item We can set $\xi_0 = 2 \,g_{3/2}\, q$ and $\xi_1 = 0$ (the factor $2$ is introduced just to canonically normalize the vector fields in the vacuum). The resulting scalar potential admits a critical point at $z=0$ and vacuum energy $V_{AdS}= -12 g_{3/2}^2q^2 $, which saturates \eqref{VAdSg2}. As a consequence, this example illustrates our analysis on scale separation in anti-de Sitter vacua: the magnetic WGC implies that the cut-off of the theory is of order of the anti-de Sitter radius, meaning that there is no separation of scales.

\item We can set $\xi_0 = 0$ and $\xi_1 = 2\, g_{3/2}\,q$, which leads to a scalar potential with critical point at $z=0$ and vacuum energy $V_{dS}= 4g_{3/2}^2q^2$, saturating \eqref{VdS>=g2}. 
This critical point is de Sitter, albeit unstable, and it illustrates our general analysis for de Sitter vacua with vanishing gravitino mass. Indeed, one can check that the matrix \eqref{gravmassmatrix} is $S_{AB} = 0$ at $z=0$. As a consequence, when applying the magnetic WGC we have that the ultraviolet cut-off of the effective description is of order of the Hubble scale and thus the model is not protected against corrections.

\end{itemize}

\paragraph{Stable de Sitter vacua.} 
Less trivial examples in which the argument against de Sitter critical points can be applied exist as well. Remarkably, as shown in \cite{Cribiori:2020use,DallAgata:2021nnr}, all known stable de Sitter vacua of $\mathcal{N}=2$ supergravity have a vanishing gravitino mass and thus are in tension with the WGC. Among those, we recall the models in \cite{Fre:2002pd,DallAgata:2021nnr}. For illustrative purposes, let us review a specific model introduced in \cite{DallAgata:2021nnr}, containing three vector multiplets and one hyper multiplet, with the gauging of an SO$(2,1)\times$U$(1)$ isometry.

The scalar manifolds are 
\begin{equation}
		{\cal S} = \left[\frac{{\rm{SU}(1,1)}}{\rm{U(1)}}\right]^3,  \qquad \qquad {\cal Q} = \frac{{\rm{SU}(2,1)}}{\rm{SU(2)}\times\rm{U(1)}}\,. 
\end{equation}
As for the special K\"ahler geometry, it might be convenient to start from a prepotential 
\begin{equation}
F(X)=\sqrt{((X^0)^2+(X^1)^2)((X^2)^2+(X^3)^2)} \, ,
\end{equation}
from which one can calculate the symplectic sections and parametrize them in terms of the Calabi-Visentini coordinates $z^i=\{S,y_0,y_1\}$; notice that they differ from the normal coordinates \eqref{normcoord} . However, the isometries to be gauged would then mix electric and magnetic sections. In order to avoid this complication, we can rotate our original symplectic frame by an angle $\phi$ and work with the rotated sections
\begin{equation}
\left(
\begin{array}{c}
X^\Lambda\\
F_\Lambda
\end{array}
\right) = \begin{pmatrix}  \frac{1}{2}( 1 + y_0^2 + y_1^2 ) \\[1mm] \frac{i}{2} (1 - y_0^2 - y_1^2) \\[1mm] y_0 \\[1mm] y_1 (\cos~ \phi - S~ \sin~ \phi) \\[1mm] 
 \frac{1}{2}S ( 1 + y_0^2 + y_1^2 ) \\[1mm] \frac{i}{2} S (1 - y_0^2 - y_1^2) \\[1mm] - S y_0 \\[1mm] - y_1 ( S~ \cos~ \phi + \sin~ \phi) \end{pmatrix} \ . 
\end{equation}
Here, $\phi$ is an example of so called de Roo--Wagemans symplectic angles. In this new frame, in which we will work from now on, the K\"ahler potential takes the form
\begin{equation}
K=-\log \left(-{\rm Im}S\left( 1 - 2|y_0|^2 - 2|y_1|^2 + |y_0^2 + y_1^2|^2 \right)\right)
\end{equation}
and the ${\rm SO}(2,1)$ isometry is generated by the Killing vectors
\begin{equation}
k_\Lambda^x = e_0 \left(k_0^i, k_1^i, k_2^i, 0\right)^T,
\end{equation}
with
\begin{align}
k_0^i = \begin{pmatrix} 0 \\[1mm] -\frac{i}{2} (1 + y_0^2 - y_1^2) \\[1mm] - i y_0 y_1 \end{pmatrix} \ , \quad
k_1^i = \begin{pmatrix} 0 \\[1mm] \frac{1}{2}( 1 - y_0^2 - y_1^2) \\[1mm] - y_0 y_1 \end{pmatrix} \ , \quad 
k_2^i = \begin{pmatrix} 0 \\[1mm] i y_0 \\[1mm] i y_1 \end{pmatrix}.
\end{align}
Here, $e_0$ plays the role of the SO$(2,1)$ coupling and the vector fields participating to the gauging are $A^0_\mu$, $A^1_\mu$ and $A^2_\mu$.

As for the quaternionic geometry, we take that of the universal hyper multiplet, see e.g.~\cite{Ceresole:2001wi}, 
parametrized by the scalar fields $q^u = \{\rho, \sigma, \theta, \tau\}$, with metric 
\begin{align}
ds^2 = h_{uv} dq^u dq^v = \frac{d\rho^2}{2 \rho^2} + \frac{1}{2 \rho^2} (d\sigma - 2 \tau d\theta + 2 \theta d \tau)^2 + \frac{2}{\rho} (d\theta^2 + d \tau^2) \ .
\end{align}
On the manifold $Q$, we gauge a compact U$(1)$ symmetry generated by the Killing vector 
\begin{equation}
k_\Lambda^u=e_1(0,0,0,k_H^u)^T,
\end{equation}
with
\begin{align}
k_H^u = \begin{pmatrix} 4 \rho \tau \\[1mm] 2 \theta + 2 \sigma \tau + 2 \rho \theta + 2 \theta(\theta^2 + \tau^2) \\[1mm] 4 \theta \tau - \sigma \\[1mm]  1 - \rho - 3 \theta^2 + \tau^2 \end{pmatrix} \ .
\end{align}
The gauging is performed with $A^3_\mu$ and $e_1$ is the corresponding coupling.

The above ingredients define the model completely and we refer the reader to \cite{DallAgata:2021nnr} for more details. By direct inspection, one can identify a critical point of the scalar potential at
\begin{equation}
S = \cot \phi - \frac{i}{4} \left|\frac{e_0}{e_1 \sin \phi} \right|,\qquad \rho = 1,\qquad y_0 = y_1 = \sigma = \theta = \tau = 0.
\end{equation}
At this point, the cosmological constant is positive (semi-)definite, $ {\cal V} = 4 | e_0 e_1 \sin \phi|$, thus giving an Hubble scale 
\begin{equation}	
		H = \sqrt{\frac{4}{3} \left|e_0 e_1 \sin \phi\right|} \ ,
\end{equation} 
while the mass matrix of the gravitini is identically vanishing. One can also check that the U$(1)$ Killing vector $k_H^u$ vanishes as well at the critical point and thus the vacuum preserves an abelian factor of the original gauge group, with respect to which we can apply the WGC.
Since at the same point the gauge kinetic matrix is diagonal
\begin{equation}
\left({\rm Im}\mathcal{N}^{-1}\right)^{\Lambda \Sigma} = - \frac{1}{2} \sin \phi \,\text{diag}\left( 4 \frac{e_1}{e_0} , 4 \frac{e_1}{e_0}, 4 \frac{e_1}{ e_0} ,\frac{ e_0 }{ e_1} \right)
\end{equation}
and the abelian gravitino charge is $q=2e_1$, we find the gravitino gauge coupling
\begin{equation}
 g_{3/2} = \sqrt{2|e_0 e_1\sin\phi|},
\end{equation}
to be identified with the magnetic WGC cutoff.
Thus, the Hubble scale is of order of the cutoff dictated by the magnetic WGC, $g_{3/2}$, and this model is a particular case of our general analysis. 
Note that the mass spectrum of the scalar fluctuations around the critical point includes two zero-modes, corresponding to the goldstone modes eaten by the two broken non-compact SO$(2,1)$ isometries,  whereas the rest of the spectrum is positive definite making the critical point perturbatively stable.

\paragraph{Other scenarios and open questions.} 
Besides the works we discussed above, we would like to briefly recall also \cite{Catino:2013syn}, in which the existence of stable de Sitter vacua in $\mathcal{N}=2$ supergravity with both (one) vector and (one) hyper multiplets is investigated as well. There, a fairly model-independent analysis is performed and general conditions to find stable de Sitter vacua are specified. Crucially, among those conditions, one finds a non-vanishing gravitino mass. As a consequence, the argument given above relying on the WGC is evaded and these vacua, if they exist, might in principle not be in the swampland. However, existence is not proven in \cite{Catino:2013syn}, and while a generic scenario is provided, concrete models are not. For this reason, we believe that this may be in principle a promising direction for future research. In particular, it would be important to (dis)prove the existence of explicit models in the scenario proposed by \cite{Catino:2013syn}.

\subsection{WGC and Festina Lente} 
\label{sec:WGCandFL}

The gravitino charge, gauge coupling and mass have played a central role in our discussion of de Sitter critical points in gauged supergravity.
In particular, whenever the gravitino had vanishing mass but non-vanishing charge, the de Sitter critical point turned out to be in tension with the WGC. 
It turns out that there exists a swampland bound enforcing such a condition not only for the gravitino but for all charged fields in the low energy effective description. 
In particular, it should hold that on a (quasi-)de Sitter background with Hubble scale $H$ and with an abelian gauge field \cite{Montero:2019ekk} 
\begin{equation}
\label{FLbound}
m^2 \gtrsim g q H M_P\,, 
\end{equation}
for every charged particle in the theory with mass $m$, charge $q$ and gauge coupling $g$. This is the so called ``Festina Lente'' bound introduced in \cite{Montero:2019ekk}  and explored further e.g.~in \cite{Montero:2021otb,Lee:2021cor,Ban:2022jgm,Guidetti:2022xct,Mohseni:2022ftn}.

By studying de Sitter critical points in gauged supergravity,  we noticed systematically that when the condition \eqref{FLbound} is violated for the gravitino, then the de Sitter background is in tension with the magnetic WGC. Therefore it seems that supergravity itself serves as a link between the WGC and the Festina Lente bound and we have explicitly seen this link at work via the gravitino. 
Notably, another condition that can be deduced from \eqref{FLbound} is \cite{Montero:2021otb}
\begin{equation}
g q M_P \gtrsim H \,, 
\end{equation}
for every charged particle in the theory. This is exactly the condition that the analyzed de Sitter critical points we studied do not satisfy, for they rather satisfy the opposite inequality \eqref{VdS>=g2}.

To summarize, our discussion suggests the possible existence of some sort of analogy
\begin{equation}
\nn
\text{Supergravity + magnetic WGC} \quad \Longleftrightarrow \quad \text{Supergravity + Festina Lente} 
\end{equation}
and one could wonder what new swampland conditions or bounds can be deduced for (anti-)de Sitter backgrounds by further applying the Festina Lente bound or the magnetic WGC on supergravity. We will come to this point later on when discussing more speculative conjectures that originate from the main course of our analysis.

\subsection{WGC and scalar fields}

Besides applying the WGC in its original formulation, another fruitful direction is to consider conjectures equivalent to (or perhaps re-formulations of) the WGC in the presence of light mediator scalar fields $\phi^i$. 
Since they can contribute as an additional attractive force felt by the particle with the largest  charge-to-mass ratio in the theory, the requirement that the total repulsive force should be stronger than the attractive one, which leads to the WGC, is modified to \cite{Palti:2017elp} 
\begin{equation}
\label{WGC+S}
2 g^2 q^2 \geq m^2 + G^{ij} Y_i Y_j. 
\end{equation}
The crucial step is to assume the mass of the WGC particle to be field dependent so that there is a Yukawa coupling $Y_i=\partial_i m(\phi)$ entering explicitly the above bound. Here, $G_{ij}$ is the field space metric for the light scalars and we are assuming only one electric charge for simplicity. We refer the reader to \cite{Palti:2017elp} for more details and generalizations. 
This condition was further studied in \cite{Lust:2017wrl} where, among other things, it was noticed that if a WGC particle couples almost equally to gauge and light scalar fields, namely $2g^2 q^2 \sim G^{ij}Y_i Y_J$, then its mass must be light, $m^2 \sim 0$, as a consequence of \eqref{WGC+S}. Notice that this holds even if the magnetic WGC cutoff $2g^2 q^2 \sim \Lambda_{UV}$ is in principle high. As such, this is an example of UV/IR mixing, for the UV consistency imposes the existence of a very light particle in the IR.

Another related conjecture holds that, even in the absence of electromagnetic interactions, one should have at least one state with the property \cite{Palti:2017elp} 
\begin{equation}
\label{SWGC} 
G^{ij} Y_i Y_j > m^2  \,, 
\end{equation}
which was dubbed scalar weak gravity conjecture in \cite{Palti:2019pca}. Notice that this time one has a strict inequality. An interesting aspect of this conjecture, which resonates with our discussion here, is that the condition \eqref{SWGC} is satisfied by the BPS states in four-dimensional $\mathcal{N}=2$ supergravity theories. This follows directly from the special geometry relation $|D_iZ|^2 - |Z|^2 =\frac12 Q^T M(F) Q$, see e.g.~formula (57) and (58) in \cite{Ceresole:1995ca}, where $Z$ is the central charge of the $\mathcal{N}=2$ algebra. Therefore, \eqref{SWGC} is satisfied in compactifications of Type II string theory on Calabi--Yau threefolds by construction.

A possible extension of \eqref{SWGC} was proposed in \cite{Gonzalo:2019gjp} in the framework of a general field theory of scalars coupled to gravity.  The idea was to impose that scalar self-interactions should always be stronger than gravity, now for any value of the scalar field in the theory. 
Such a strong bound results in the condition \cite{Gonzalo:2019gjp} 
\begin{equation}
2 (V''')^2 - V'' V '''' \geq \frac{(V'')^2}{M_P^2} \,,  
\end{equation}
where prime refers to the derivative with respect to the probe scalar field, e.g.~$V' = \frac{\partial V }{ \partial \phi}$ for a real probe scalar $\phi$. Staying outside the explicit domain of supergravity theories, in \cite{Benakli:2020pkm} the impact of the scalar weak gravity conjectures was studied for various cosmological scenarios.

Inspiration from supergravity in deriving and refining swampland bounds was further exploited in \cite{Palti:2017elp}. It is known that the central charge of the $\mathcal{N}=2$ supersymmetry algebra satisfy a specific identity that depends on the derivatives with respect to the special K\"ahler scalar fields. Extrapolating this relation into a swampland conjecture, one can obtain an inequality of the form \cite{Palti:2017elp} 
\begin{equation}
n \, m^2 + G^{ij} Y_i Y_j \leq \frac12 D_i \partial_j ( m^2 ) \,, 
\end{equation}
where now $n$ is the number of scalar fields coupled to the WGC state.  This conjecture was further studied in \cite{DallAgata:2020ino}, especially in relation to the identities that the central charges of $\mathcal{N}\geq 2$ supergravity satisfy.

\section{Introducing strings and membranes}

In the previous section, swampland conjectures have been applied directly to supergravity effective theories with particles as matter content. 
A possible generalization is to supplement the theory with additional ingredients and then study swampland conjectures in such an enriched setup. 
These ingredients can be extended objects, as for example string or membranes, and they are inserted directly in the four-dimensional theory. 
Such a step can be motivated by the completeness hypothesis \cite{Polchinski:2003bq}.
This approach has been recently pursued in \cite{Lanza:2020qmt,Lanza:2021udy,Martucci:2022krl,Marchesano:2022avb,Marchesano:2022axe} (see \cite{Katz:2020ewz} for a similar work in five dimensions), based on the manifest supersymmetric formalism developed in \cite{Farakos:2017jme,Bandos:2018gjp}, but the history of supergravity coupled to extended objects dates back to \cite{Binetruy:1996xw,Ovrut:1997ur}.

In this section, we briefly review these recent developments, with particular focus on \cite{Martucci:2022krl}, since they provide a somehow complementary approach to the one we presented above. 
Any misconception in the presentation of the work by other authors is of course ours.

\subsection{String and membranes in the perturbative regime of the effective theory}
\label{sec:EFTstrings}

Besides particles, in four dimensions one can have strings and membranes which are co-dimension two and one objects respectively. When introduced in the effective description as probes, one can derive consistency conditions for the whole theory by demanding consistency of the probe. This approach has been followed in five dimensions in \cite{Katz:2020ewz} and more recently in four dimensional $\mathcal{N}=1$ supergravity in \cite{Martucci:2022krl}. Before entering the discussion of \cite{Martucci:2022krl}, we would like to review how strings and membranes can be introduced into the setup of minimal supergravity in four dimensions.

First, we would like to outline the regime we will be working in. One can define a perturbative regime of the effective theory by looking at the gauge sector and demanding it to be weakly coupled. In the same regime, non-perturbative contributions to the superpotential are suppressed. This is described below.

In the four-dimensional effective theories of interest there are axions $a^i$, with fixed periodicity
\begin{equation}
a^i \simeq a^i +1,
\end{equation}
and which, together with saxions $s^i$, form the lowest components of $\mathcal{N}=1$ chiral multiplets
\begin{equation}
t^i = a^i + i s^i.
\end{equation}
Besides, one can also have spectator chiral multiplets $\phi$ such that the full set of scalars $z=\{t^i, \phi\}$ parametrizes a K\"ahler-Hodge field space. From a bottom-up perspective, axions are associated to continuous shift isometries, $a^i \to a^i + c^i$, which are usually broken to discrete transformations by means of BPS instanton effects,
\begin{equation}
W_{inst} \simeq e^{2\pi m_i t^i},
\end{equation}
with $m_i$ the BPS instanton charges. The perturbative regime we will be working in is defined by the condition
\begin{equation}
\left| e^{2 \pi m_i t^i} \right| = \left| e^{-2\pi m_i s^i}\right| \ll 1,
\end{equation}
guaranteeing that the continuous shift symmetry of the axions is approximately preserved, while the K\"ahler potential depends only on the saxions
\begin{equation}
K = K(s).
\end{equation}
This regime corresponds to weak coupling in the gauge sector. Without loss of generality, we consider a gauge group $G = \Pi_A {\rm U}(1)_A \, \times \Pi_I G_I$, where $G_I$ are simple groups. The couplings of vector to scalar fields are given by the kinetic terms\footnote{With respect to section \ref{sec:N=1rev}, we rescale the gauge kinetic function by a factor of $1/2\pi$ for abelian groups and $1/4\pi$ for $G_I$, to match with the conventions of \cite{Martucci:2022krl}. Besides, we split the vector field indices as $\Lambda = (A,I)$ in accordance with the splitting of the gauge group $G$.}
\begin{equation}
\label{kinvecpert1}
\begin{aligned}
-\frac{1}{4\pi}&\int\left[ ({\rm Im} f)^{AB} F_A\wedge * F_B + ({\rm Re}f)^{AB} F_A \wedge F_B\right] \\
&\qquad \qquad -\frac{1}{8\pi} \int \left[{\rm Im}f^I {\rm tr}(F \wedge *F)_I + {\rm Re} f^I {\rm tr}(F\wedge F)_I\right].
\end{aligned}
\end{equation}
We assume that we are allowed to expand
\begin{align}
f^{AB}(t,\phi) &= C_i^{AB}t^i+\Delta f^{AB}(\phi)+\dots,\\
f^I(t,\phi) &= C_i^I t^i + \Delta f^I(\phi) +\dots,
\end{align}
where $C_i^{AB}$, $C_i^I$ are constants and dots stand for terms which will be exponentially suppressed. Then, we see that the perturbative regime corresponds to ${\rm Im}f^{AB}, {\rm Im}f^I\gg 1$, which is indeed weak gauge coupling, and the dominant terms in \eqref{kinvecpert1} are
\begin{equation}
\label{axcoup1}
-\frac{1}{4\pi}C_i^{AB}\int\left[ s^i F_A\wedge * F_B + a^i F_A \wedge F_B\right] -\frac{1}{8\pi} C_i^I\int \left[s^i {\rm tr}(F \wedge *F)_I + a^i {\rm tr}(F\wedge F)_I\right]+\dots.
\end{equation}
The very form of these couplings will be a working assumption for the discussion of quantum gravity constraints in the next sections. Besides, it is expected that $\Delta f^{AB}(\phi)$ and $\Delta f^{I}(\phi)$ are order one contributions for generic values of $\phi$, so that one has to require $C_{i}^{AB}s^i \geq 0$ and $C_i^I s^i \geq 0$ to avoid ghosts in this regime. Compatibility of gauge instanton corrections with axion periodicity requires in addition that these coefficients are quantized as $C_i^{AB},C_i^I \in \mathbb{Z}$.

As we have seen in section \ref{sec:N=1rev}, we can also consider higher curvature corrections to the supergravity action, such as those in formula \eqref{R2corrections}. Similarly to the gauge kinetic function, we can expand $\tilde f = \tilde f(t,\phi)$ as
\begin{equation}
\tilde f(t,\phi) = \tilde C_i t^i + \Delta \tilde f(\phi) +\dots,
\end{equation}
where again dots stand for  non-perturbative terms which are exponentially suppressed in the perturbative regime. Thus, from \eqref{R2corrections} we read the coupling
\begin{equation}
\label{axcoup2}
-\frac{1}{96\pi} \tilde C_i \int \left[s^i {\rm tr}\left(R \wedge * R\right)+a^i {\rm tr}\left(R \wedge R\right)\right],
\end{equation}
where the quantization condition $2 \tilde C_i \in \mathbb{Z}$ is needed not to break the axion periodicity by possible gravitational instantons.

Having specified the main features of the perturbative regime we are considering, we want now to introduce (probe) strings and membranes in the setup. We mainly focus on the former, and indeed we consider a specific class of BPS axionic strings, called EFT strings for short \cite{Lanza:2020qmt,Lanza:2021udy}. 
They have the property that their backreaction on the moduli is such that close to the string core the effective theory becomes weakly coupled, namely axionic shift symmetries become exact. In this respect, EFT strings can be considered as fundamental, in the sense that they cannot be completed into smooth solitonic objects within the effective description.

More concretely, EFT strings are detected by axions, which undergo integral shifts around them,
\begin{equation}
a^i \to a^i + e^i, \qquad e^i \in \mathbb{Z}.
\end{equation}
They have a field-dependent tension
\begin{equation}
T = M_P^2 \,e^i l_i,
\end{equation}
where $l_i = - \frac12 \frac{\partial K}{\partial t^i} = -\frac12 \frac{\partial K}{\partial s^i}$ are linear multiplets dual to the saxions and $e^i$ are the charges entering the coupling with the 2-form potential $B_{2,i}$, namely $e^i\int_\mathcal{W} B_{2,i}$, where $\mathcal{W}$ is the string worldsheet. The tension is such that $\Lambda_{EFT}^2 < T < M_P^2$, which is another way to see that these strings cannot be resolved within the effective theory and as such are fundamental. Close to their core, the backreaction induces a profile for the saxions of the type \cite{Greene:1989ya,Dabholkar:1990yf}
\begin{equation}
s^i(r) = s^i_0 + \frac{e^i}{2\pi} \log \frac{r_0}{r},
\end{equation}
where $s^i_0 \equiv s^i(0)$ and $r$ is the coordinate parametrizing the distance from the core sitting at $r=0$. Thus, we see that the backreaction becomes strong approaching the core, such as $s^i \gg 1$ and thus the effective theory is in the perturbative regime.

Membranes can be introduced by generalizing the discussion on strings. In particular, in four-dimensional $\mathcal{N}=1$ supergravity a natural way to introduce membranes is to first trade scalar auxiliary fields for (hodge duals of) gauge four-form field strengths. Remarkably, this step can be performed in a manifestly supersymmetric manner \cite{Gates:1980ay,Binetruy:1996xw,Ovrut:1997ur,Binetruy:2000zx,Farakos:2017jme,Bandos:2018gjp, Cribiori:2020wch} (see \cite{Cribiori:2018jjh} for an analysis in global $\mathcal{N}=2$ supersymmetry) and we will come back to it in section \ref{sec:goldevap}.\footnote{In the matter-coupled old minimal formulation of supergravity there can be mainly three kinds of scalar auxiliary fields: a complex scalar from the gravity multiplet, one complex scalar from each chiral multiplet and one real scalar from each vector multiplet. The supersymmetric procedure to replace them with gauge four-form field strengths has been outlined in \cite{Ovrut:1997ur} for the first kind of auxiliary field, then extended in \cite{Gates:1980ay,Binetruy:1996xw,Binetruy:2000zx,Farakos:2017jme,Bandos:2018gjp} to auxiliary field of chiral multiplets, and more recently in \cite{Cribiori:2020wch} to the auxiliary field of vector multiplets. The setup in \cite{Cribiori:2020wch} allows in fact for a dynamical generation of the Einstein-Hilbert kinetic term from a scale-invariant action. As a result, the Planck mass and the cosmological constant are scanning variables and membranes can interpolate between vacua with different values of these quantities. As such, the model in \cite{Cribiori:2020wch} can be seen as an explicit embedding into $\mathcal{N}=1$ supergravity of a mechanism to lower dynamically the value of the cosmological constant and of the Planck mass. More recently, this mechanism has been further considered in the subsequent works \cite{Kaloper:2022oqv,Kaloper:2022utc,Kaloper:2022jpv}.} Then, membranes are the objects coupling to these newly introduced gauge three-form fields $C^a_3$, such that their local action contains a term $q_a \int_{\mathcal{W}_{3}} C_3^a$, with $\mathcal{W}_3$ the membrane worldvolume. They can act as domain walls separating, say, two different $\mathcal{N}=1$ flux vacua. When crossing the wall, certain fluxes jump by $q_a$ units and in turn the value of the cosmological constant is changed. 

An interesting question is how these strings and membranes behave with respect to the WGC. This has been investigated in \cite{Lanza:2020qmt,Lanza:2021udy} to which we refer the reader for more details. Here, we would just like to point out that a similar question is behind the works reviewed in section \ref{sec:WGCvs}. In fact, one of the results of \cite{Lanza:2020qmt} is that imposing the WGC for fundamental membranes leads to a runaway contribution to the scalar potential forbidding de Sitter critical points.

\subsection{Anomaly inflow on EFT strings}

The idea of \cite{Martucci:2022krl} is to derive constraints on four-dimensional $\mathcal{N}=1$ supergravity by asking cancellation of anomalies on the probe EFT string with an anomaly inflow from the bulk theory. In practice, one has to calculate the possible anomalies on the string worldsheet theory and in the bulk supergravity theory separately, and then ask that their combination cancels.
Following this strategy, one can constrain the bulk supergravity theory itself by asking consistency of the probe string.

As for the bulk theory, the axionic couplings produce an anomaly inflow to the EFT string. In the setup under investigation, we can recast the axionic couplings from \eqref{axcoup1} and \eqref{axcoup2} as
\begin{equation}
2 \pi \int a^i I_{4,i} = -2 \pi \int h^i_1 \wedge I_{3,i}^{(0)},
\end{equation}
where we introduced the globally defined one-forms $h^i_1 = da^i$ and we set
\begin{equation}
I_{4,i} = -\frac{1}{8\pi^2} C_i^{AB} F_A \wedge F_B - \frac{1}{16\pi^2} C_i^I {\rm tr} (F \wedge F)_I - \frac{1}{192 \pi^2} {\rm tr} (R \wedge R)_I\, .
\end{equation}
This object is an example of invariant polynomial, see e.g.~the standard textbook \cite{Nakahara:2003nw}. According to Chern-Weil theorem, invariant polynomials are closed (and do not dependent on the choice of a particular connection). Hence, by Poincar\'e lemma they are locally exact and one can write
\begin{equation}
I_{4,i} = d I^{(0)}_{3,i}.
\end{equation}
The object $I^{(0)}_{3,i}$ is called transgression in general, and Chern-Simons form in our particular case. Asking that it describes a theory on its own, invariant under general gauge and local Lorentz transformations, amounts to ask that
\begin{equation}
\delta I_{3,i}^{(0)} = d I_{2,i}^{(1)},
\end{equation}
for some two-form $I_{2,i}^{(1)}$. The latter is the anomaly inflow induced by the axionic couplings on the worldsheet of the string.  
Indeed, recalling that the axion field strengths $h_1^i$ in the presence of a string satisfy the Bianchi identity
\begin{equation}
\label{dhBianchi}
dh^i = e^i \delta_2 (\mathcal{W}),
\end{equation}
we can calculate the variation of the bulk axionic couplings
\begin{equation}
\begin{aligned}
\delta S_{bulk}&=\delta \left(2 \pi \int a^i I_{4,i}\right) = - 2 \pi \int h^i_1 \wedge \delta I^{(0)}_{3,i} \\&=-2 \pi e^i \int \delta_2(\mathcal{W}) \wedge I_{2,i}^{(1)}  = -2 \pi e^i \int_{\mathcal{W}} I_{2,i}^{(1)} .
\end{aligned}
\end{equation}
Since this is non-vanishing, there is an anomaly which must be cancelled by an analogous contribution from the worldsheet string action, namely we should find
\begin{equation}
\delta S_{string} = 2\pi e^i \int_{\mathcal{W}} I_{2,i}^{(1)}.
\end{equation}
In other words, by consistency  the object $I_4^{WS}=e^iI_{4,i}$ is required to be the anomaly polynomial of the worldsheet theory.

Before proceeding further, we need two perform two different steps. First, we want to rewrite $I_4^{WS}$ in a more convenient form. To this purpose, we split the contribution of the bulk first Pontryagin class $p_1(M) = -\frac{1}{8\pi^2}{\rm tr} \left(R \wedge R\right)$ into tangent and normal contributions with respect to the worldsheet. This can be done as\footnote{One needs to use $TM|_\mathcal{W} = T\mathcal{W} \oplus N_{\mathcal{W}}$, where $N_{\mathcal{W}}$ is the normal bundle of $\mathcal{W}$, together with Whitney sum formula implying $p_1(TM|_{\mathcal{W}}) = p_1(T\mathcal{W})+p_1(N_{\mathcal{W}})$. Then, one uses also that in general $p_1 = -2 c_2 + c_1^2$, but $c_2(N_{\mathcal{W}})=0$ in our case.}
\begin{equation}
\label{p1split}
p_1(M) = p_1 (\mathcal{W}) + c_1(N_\mathcal{W})^2 = -\frac{1}{8\pi^2}{\rm tr}\left(R_{\mathcal{W}}\wedge R_{\mathcal{W}}\right) + \frac{1}{4\pi^2} \left(F_N \wedge F_N\right),
\end{equation}
where $F_N = dA_N$ is the field strength of the U$(1)_N$ connection induced on the normal bundle  $N_{\mathcal{W}}$ by the bulk Riemannian connection.

Second, there is an additional contribution to be taken into account. This arises by requiring that the normal bundle $N_{\mathcal{W}}$ must be trivial, similarly to the discussion in \cite{Witten:1996hc} for the normal bundle to the $NS5$-brane in type IIA string theory. Intuitively, we are considering a (probe) string in an effective theory and thus we have to glue properly the worldsheet theory (the zero-modes perturbation theory) to the bulk. In this respect, the delta function on right hand side of the Bianchi identity \eqref{dhBianchi} can be seen as a tool to keep track of the string. A more precise mathematical description is available in cohomology, where the right hand side of \eqref{dhBianchi}, and also its restriction to $\mathcal{W}$, can be seen as a two-form representative of a certain class. By combining proposition 6.24 and 6.41 of \cite{BottTu}, one finds that the pullback of this cohomology class to the string worldsheet is the Euler class of the normal bundle, namely
\begin{equation}
\delta_2(\mathcal{W})\big|_{\mathcal{W}} = \chi(N_\mathcal{W}).
\end{equation}
A natural choice for a connection on this bundle is $A_N$ and thus we can write
\begin{equation}
\chi(N_\mathcal{W}) = \frac{1}{2\pi} F_N.
\end{equation}
Then, combining with \eqref{dhBianchi}, we get a finite contribution from the delta source
\begin{equation}
\label{dhBianchi2}
dh^i\big|_{\mathcal{W}} = \frac{e^i}{2\pi}F_N.
\end{equation}
This is the precise sense in which the normal bundle to the string worldsheet is trivial in cohomology and we will take it as a defining property for EFT strings, implying in turn that $F_N = dA_N$ globally. The Bianchi identity \eqref{dhBianchi2} suggests the presence of an additional term in the effective action localized on the string and of the type
\begin{equation}
S_N = -\frac{1}{24} \hat C_i \int_{\mathcal{W}} h^i_1 \wedge A_N.
\end{equation}
Due to \eqref{dhBianchi2}, this term is anomalous under U$(1)$ gauge transformations $\delta A_N = d\lambda_N$, indeed
\begin{equation}
\delta S_N = -\frac{1}{24} \hat C_i \int_{\mathcal{W}} dh^i_1 \wedge \lambda_N = -\frac{1}{48\pi} \hat C_i e^i \int_{\mathcal{W}} F_N \lambda_N.
\end{equation}
In turn, this implies the existence of a new contribution to the anomaly polynomial, namely 
\begin{equation}
\label{I4N}
\hat I_{4,N} = \frac{\hat C_i e^i}{96\pi^2} F_N \wedge F_N.
\end{equation}
The total worldsheet anomaly polynomial is thus the sum of $e^i I_{4,i}$, written taking into account \eqref{p1split}, and the additional piece \eqref{I4N}, giving
\begin{equation}
\begin{aligned}
\label{anomalypoltot}
I_4^{WS} &= e^i I_{4,i} + \hat I_{4,N} \\
&=-\frac{C_i^{AB} e^i}{8\pi^2} F_A \wedge F_B - \frac{C_i^I e^i}{16 \pi^2}{\rm tr}\left(F \wedge F\right)_I \\&\quad - \frac{\tilde C_i e^i}{192 \pi^2}{\rm tr}\left(R_{\mathcal{W}}\wedge R_{\mathcal{W}}\right) + \frac{\tilde C_i e^i+\hat C_i e^i}{96 \pi^2} F_N \wedge F_N.
\end{aligned}
\end{equation}

\subsection{Constraints on $\mathcal{N}=1$ supergravity}

The idea is to match the anomaly polynomial \eqref{anomalypoltot} with the one arising from the string worldsheet, which can be calculated directly. It contains one contribution from gravitational and U$(1)_N$ anomalies and another from 't Hooft gauge anomalies. We discuss both of them separately below.

The first contribution has been originally calculated in \cite{Alvarez-Gaume:1983ihn} and reads 
\begin{equation}
\label{gravU1an}
I^{WS}_4 |_{grav+U(1)_N} = -\frac{n_F -n_C+n_N-1}{192 \pi^2} {\rm tr}\left(R_{\mathcal{W}}\wedge R_{\mathcal{W}}\right) + \frac{n_C -n_F+1}{32\pi^2} F_N\wedge F_N.
\end{equation}
Here, $n_F$, $n_C$ and $n_N$ are respectively the number of chiral U$(1)_A$-charged fermi multiplets, the number of chiral U$(1)_A$-charged scalar multiplets and the number of chiral U$(1)_N$-charged fermi multiplets of the $\mathcal{N}=(0,2)$ non-linear sigma-model worldsheet theory. Besides, there is also an additional universal U$(1)_A$-charged chiral multiplet to be taken into account. At a generic point of the moduli space one expects that $n_C-n_N\geq -1$, for an enhancement to $\mathcal{N}=(2,2)$ supersymmetry requires $n_F=0$ and $n_C-n_N=-1$, while otherwise it is reasonable to assume $n_C-n_N\geq 0$. We refer to \cite{Martucci:2022krl} for more details.
By matching the coefficients of \eqref{gravU1an} with those of the analogous terms in \eqref{anomalypoltot}, one finds
\begin{align}
\label{QGconstr1}
\tilde C_i e^i = n_F -n_C+n_N-1,\\
\label{QGconstr2}
\tilde C_i e^i + \hat C_i e^i = 3(n_C-n_N+1).
\end{align}
Assuming completeness of the EFT string spectrum gives then
\begin{align}
\label{QGconstr12}
&\tilde C_i e^i = n_F -n_C+n_N-1,\\
\label{QGconstr22}
&\tilde C_i e^i + \hat C_i e^i \in 3 \mathbb{Z}_{>0}.
\end{align}
These are neat examples of quantum gravity constraints on the parameters of $\mathcal{N}=1$ supergravity which are not obvious from a low energy perspective.

Next, we look at 't Hooft gauge  anomalies. They can receive contributions from the $n_N$ fermi multiplets and the $n_C$ chiral multiplets. Their anomaly polynomial will be of the form
\begin{equation}
\label{anpolgauge}
-\frac{1}{8\pi^2} \kappa^{\mathcal{A}\mathcal{B}} F_{\mathcal{A}} \wedge F_\mathcal{B},
\end{equation}
where we collected all field strengths in $F_{\mathcal{A}} = (F_A, F_{I\alpha})$, with $\alpha=1,\dots,rk(g_I)$, and we defined $\kappa^{\mathcal{A}\mathcal{B}}=\kappa^{\mathcal{A}\mathcal{B}}_F+\kappa^{\mathcal{A}\mathcal{B}}_C$, with the first term associated to fermi multiplets and the second to chiral multiplets. Tacitly, we have already decomposed ${\rm tr}(F\wedge F)_I$ on a basis of the Cartan $h_I \subset g_I$, with $g_I$ being the algebra of $G_I$. The goal is to derive an upper bound for the rank of the gauge group $G$, which is defined as
\begin{equation}
\mathrm{r}(e) = {\rm rank}\{C_i^{AB}e^i\} + \sum_{I: C^I_i e^i>0} rk(g_I).
\end{equation}
and receives a contribution from fermi multiplets,
\begin{equation}
{\rm r}_F(e) = {\rm rank}\{\kappa_F^{AB}(e)\} + \sum_{I:\kappa_F^I>0} rk(g_I) \equiv {\rm rank}\{\kappa_F^{\mathcal{A}\mathcal{B}}(e)\},
\end{equation}
and from chiral multiplets
\begin{equation}
{\rm r}_C(e) = {\rm rank}\{\kappa_C^{\mathcal{A}\mathcal{B}}(e)\}.
\end{equation}
Using that for any two matrices $M_1$, $M_2$ one has ${\rm rank}(M_1+M_2)\leq {\rm rank}(M_1)+ {\rm rank}(M_2)$, one can write
\begin{equation}
{\rm r}(e) \leq {\rm r}_F(e) + {\rm r}_C(e)
\end{equation}
and then the strategy is to derive upper bounds for ${\rm r}_F(e)$ and ${\rm r}_C(e)$ separately. The analysis is involved and must be performed with care. We do not aim at presenting a complete derivation, rather we just sketch the main steps and refer the reader to \cite{Martucci:2022krl} for more details.

We start from ${\rm r}_F(e)$, namely the contribution from fermi multiplets. Without loss of generality, we can write
\begin{equation}
\kappa_F^{\mathcal{A}\mathcal{B}} = \sum_{{\rm q} \in \text{fermi}}{\rm q}^\mathcal{A}{\rm q}^{\mathcal{B}},
\end{equation}
which is a sum of $n_F$ matrices ${\rm q}^\mathcal{A}{\rm q}^{\mathcal{B}}$ with rank either $0$ or $1$. Thus, we get that
\begin{equation}
{\rm r}_F(e) ={\rm rank}\{\kappa_F^{\mathcal{A}\mathcal{B}}(e)\} \leq n_F.
\end{equation}
Combining this with \eqref{QGconstr1} and \eqref{QGconstr2}, one arrives at the bound
\begin{equation}
\label{rkbound1}
{\rm r}_F(e) \leq n_F = \frac43 \tilde C_i e^i +\frac13 \hat C_i e^i.
\end{equation}

To give a bound on ${\rm r}_C(e)$ is more subtle. If the chiral multiplets are charged under the gauge group, their fermions contribute negatively to the anomaly polynomial and so they cannot increase the rank of the gauge algebra. The only possibility for a chiral multiplet to contribute to the upper bound on the rank is then the situation in which such multiplet has an axionic shift symmetry gauged by the four-dimensional gauge symmetry. In the present setup, this can happen for at most $n_C-n_N$ chiral multiplets. Working out the details, one finds that they contribute to the anomaly polynomial as in \eqref{anpolgauge} and one can derive that
\begin{equation}
{\rm r}_C(e) = {\rm rank}\{\kappa_C^{\mathcal{A}\mathcal{B}}(e)\} \leq 2 \left(n_C-n_N\right),
\end{equation}
where the factor $2$ arises from the fact the matrix $\kappa_C^{\mathcal{A}\mathcal{B}}(e)$ has two pieces giving the same contribution. Using \eqref{QGconstr2}, this becomes
\begin{equation}
\label{rkbound2}
{\rm r}_C(e) \leq \frac 23 \tilde C_i e^i + \frac23 \hat C_i e^i -2.
\end{equation}
Eventually, one finds that the total rank ${\rm r}(e)$ is bounded as
\begin{equation}
{\rm r}(e) \leq {\rm r}_F(e) + {\rm r}_C(e) \leq n_F + (n_C-n_N),
\end{equation}
or equivalently
\begin{equation}
\label{QGrkbound}
{\rm r}(e) \leq 2 \tilde C_i e^i + \hat C_i e^i -2.
\end{equation}
This indicates a correlation between the rank of the gauge algebra and higher curvature corrections which is not obvious a priori from a low energy perspective. 
To summarize, the quantum gravity constraints \eqref{QGconstr12}, \eqref{QGconstr22} and \eqref{QGrkbound} are among the main results of \cite{Martucci:2022krl}.

\subsubsection{Example: $D3$-brane tadpole from bottom-up}

It is illustrative to study the quantum gravity constraints reviewed above in a concrete models. 
Here, we consider an example taken from section 6.6 of \cite{Martucci:2022krl}, but we present it in a slightly different way, to show that in fact those constraints are powerful enough to derive certain string theory tadpole cancellation conditions from an (almost) exclusively bottom-up perspective. Several additional and interesting examples are discussed at length in \cite{Martucci:2022krl}, to which we most definitely refer the reader.

The setup is type IIB string theory on an orientifold of a Calabi-Yau threefold with $O3$-planes, $D3$- and $D7$-branes. The $D3/O3$ sources are spacetime filling while the $D7$-branes wrap the full Calabi-Yau and can be seen as EFT strings in the non-compact space. As it is known, in the absence of $O7$-planes, but also of $D7$-branes wrapping four-cycles and of fluxes, the integrated $D3$-brane tadpole reduces to
\begin{equation}
\label{D3tad}
n_{D3} = \frac{n_{O3}}{4},
\end{equation}
where $n_{D3}$ and $n_{O3}$ are the number of $D3$-branes and $O3$-planes respectively. The goal is to derive this tadpole from the constraints discussed above (and with one additional assumption).

To start, we have to identify the various couplings of the effective $\mathcal{N}=1$ supergravity theory and match them with the microscopic quantities. There is one relevant scalar field, namely the axio-dilaton
\begin{equation}
t = a+is = C_0^{RR} + i e^{-\phi}.
\end{equation}
Furthermore, we assume the $D3$-branes to be non-coincident, so that they are described by the abelian couplings 
\begin{equation}
-\frac{1}{4\pi} \sum_{A=1}^{n_{D3}} s F_A \wedge* F_A.
\end{equation}
With respect to \eqref{axcoup1}, we thus have $C^{AB} = \delta^{AB}$ and $C^I =0$.
Higher curvature couplings are known from string theory and they can be read off from the Chern-Simons action of the sources. For the $D3$-branes, one has the $A$-roof genus,
\begin{equation}
-2\pi n_{D3} \int C_0^{RR} \sqrt{\hat A(R)} = \frac{2\pi n_{D3}}{48} \int a \,p_1(M)+\dots,
\end{equation}
while for the $O3$-planes one has the Hirzebruch $L$-polynomial
\begin{equation}
\frac{2\pi n_{O3}}{4}\int C_0^{RR}\sqrt{L\left(R/4\right)} = \frac{2\pi}{4}\frac{n_{O3}}{96}\int a\, p_1(M)+\dots .
\end{equation}
Here, $p_1(M)=-\frac{1}{8\pi^2}{\rm tr}R\wedge  R$ is the first Pontryagin class of the tangent bundle of the spacetime $M$. Putting everything together, we have the axionic coupling
\begin{equation}
\frac{2\pi (8n_{D3}+n_{O3})}{4 \cdot 96} \int a\,p_1(M) = -\frac{8 n_{D3}+n_{O3}}{16 \cdot 96 \pi} \int a\, {\rm tr}(R \wedge R) \equiv -\frac{1}{96 \pi}\tilde C \int a \,{\rm tr}(R \wedge R),
\end{equation}
from which we identify the macroscopic parameter $\tilde C$ in terms of microscopic data as
\begin{equation}
\tilde C = \frac{1}{16}(8n_{D3}+n_{O3}).
\end{equation}
Here and in the following we absorb the charge $e$ into $\tilde C$, or equivalently we set it to one.

Now, let us look at the bound \eqref{QGrkbound} on the gauge group. Since $\hat C_i e^i=0$ in this example, we have
\begin{equation}
{\rm r}(e) \leq 2 \tilde C-2 = \frac{8 n_{D3}+n_{O3}}{16}-2.
\end{equation}
That the right hand side is an integer follows from the quantization condition $\tilde C_i e^i \in \mathbb{Z}$ discussed in section \ref{sec:EFTstrings}. This leads to interesting constraints, such as that $n_{O3}$ must be a multiple of sixteen, as pointed out in \cite{Martucci:2022krl}. However, we proceed below following a slightly different path with respect to \cite{Martucci:2022krl}. Let us assume that chiral multiplets do not contribute to the rank ${\rm r}(e)$, namely ${\rm r}_C(e)=0$. Thus, we have
\begin{equation}
{\rm r}(e) \leq {\rm r}_F(e) \leq n_F = \frac43 \tilde C_i e^i = \frac{1}{12} \left(8n_{D3}+n_{O3}\right).
\end{equation}
Given that the gauge group is associated to $D3$-branes only, the rank cannot be bigger than $n_{D3}$ and then
\begin{equation}
{\rm r}(e) \leq \frac{1}{12} \left(8n_{D3}+n_{O3}\right) \equiv n_{D3}.
\end{equation}
From this, one finds \eqref{D3tad} with precisely the same coefficients. We have shown that, modulo the assumption ${\rm r}_C(e)=0$, the quantum gravity constraints  \eqref{QGconstr12}, \eqref{QGconstr22} and \eqref{QGrkbound} succeeded in reproducing string theory information from a bottom-up perspective. It would be interesting to see if a similar logic can be pursued also in other setups.

\subsection{Goldstino evaporation} 
\label{sec:goldevap}

In this section, we review another application of supergravity in the presence of membranes which has been discussed in \cite{Farakos:2020wfc}. We show that this formalism allows for an explicit description of a non-smooth transition between a non-supersymmetric vacuum with non-linearly realized supersymmetry and a supersymmetric one. The transition happens due to the presence of membranes and it can have important consequences on the fate of de Sitter vacua in $\mathcal{N}=1$ supergravity.

The archetypal model for supersymmetry breaking is the Volkov--Akulov model \cite{Volkov:1973ix} 
(for the supergravity coupling see e.g.~\cite{Lindstrom:1979kq,Dudas:2015eha,Bergshoeff:2015tra,DallAgata:2016syy,Cribiori:2016qif}), 
which has a single goldstino fermion $G$ whose self-interactions are governed by the lagrangian 
\begin{equation}
{\cal L}_{VA}= -f^2 - \frac12 \overline G P_L \partial \!\! \! / G 
+ \frac{1}{4 f^2} \overline G^2 \partial^2 G^2 
-\frac{1}{16 f^6} G^2 \overline G^2 \partial^2 G^2 \partial^2 \overline G^2 \, , 
\end{equation}
where $\sqrt f$ is the supersymmetry breaking scale and we use the notation $G^2=\overline G P_L G$. 
Once gauge three-forms $C_3$ are introduced in supersymmetry and supergravity, as briefly reviewed at the end of section \ref{sec:EFTstrings}, one also has the option to support the breaking of supersymmetry with four-form flux, that is one can have 
\begin{equation}
f = \star dC_3 \, , 
\end{equation}
for a real three-form $C_3$. 
The way to achieve this self-consistently with non-linear supersymmetry has been analyzed in \cite{Farakos:2016hly,Farakos:2020wfc}, to which we refer the reader for more details. 
What we want to review instead is how the systems with inherently non-linear realizations of supersymmetry can decay to systems where supersymmetry is restored and the goldstino {\it evaporates}, as depicted in \cite{Farakos:2020wfc}. More broadly, this resonates with the expectation that non-supersymmetric vacua should decay.\footnote{Another way that non-supersymmetric vacua could decay, is related to {\it goldstino condensation} \cite{DallAgata:2022abm,Kallosh:2022fsc,Farakos:2022jcl}, or the equivalent effect from the gravitino perspective \cite{Alexandre:2013iva,Alexandre:2014lla}.}

To describe the goldstino evaporation, we have first to exchange one of the two real components of the complex auxiliary field $F$ of the nilpotent goldstino multiplet with the gauge three-form. Thus, we set
\begin{equation}
F =  \mathcal{F} + i \partial_\mu C^\mu = \mathcal{F} + i\star dC_3\, , 
\end{equation}
where $\mathcal{F}$ is a real scalar and the vector $C_\mu$ is defined via $C_3 = \frac{1}{3!} dx^c \wedge dx^b \wedge dx^a C_{abc}=\frac{1}{3!} dx^c \wedge dx^b \wedge dx^a \epsilon_{abcd} C^d$. 
In the simple setup studied in \cite{Farakos:2020wfc} we have the vacuum-expectation-values 
\begin{equation}
\langle \mathcal{F} \rangle = 0, \qquad \langle \star dC_3 \rangle = {\rm n} \,, 
\end{equation}
which are telling us that supersymmetry is broken only due to the four-form flux. 
Once the three-form has been introduced, one can couple it to a (super-)membrane, 
leading to an overall bosonic sector of the form 
\begin{equation}
\begin{aligned}
S_{\text{bos}} = & \int d^4 x \, \left[ \mathcal{F}^2 + \left(  \partial_\mu C^\mu \right)^2 
- 2 \partial_\mu \left( C^\mu \partial_\nu C^\nu \right) \right] \\
& - \frac{c}{4 \pi}  \int_{\mathcal{W}_3} d^3 \zeta \, \sqrt{- \text{det} \, (\eta_{ab} \partial_i x^a(\zeta) \partial_j x^b(\zeta))}  
- \frac{\mu}{4 \pi}  \int_{\mathcal{W}_3} C_3 \, , 
\end{aligned}
\end{equation}
where $\mathcal{W}_3$ is the membrane worldvolume, which is parametrized by coordinates $\zeta^i$, with $i=0,1,2$, while $c$ is the tension and $\mu$ the charge under the three-form. 
Assuming completeness of the spectrum, one could argue that the existence of charged membranes is natural \cite{Polchinski:2003bq}. 
We see immediately that by varying $\mathcal{F}$ we get $\mathcal{F}=0$. 
Varying instead the action with respect to the three-form we find that 
\begin{equation}
\label{Ceom}
\partial_\mu \partial_\nu C^\nu = - \frac{\mu}{48 \pi} \int_{\mathcal{M}_3} d^3 \zeta \, \epsilon^{kji} \partial_k x^\lambda(\zeta) \partial_j x^\sigma(\zeta) \partial_i x^\nu(\zeta) \epsilon_{\nu \sigma \lambda \mu} \delta^{(4)} (x-x(\zeta)) \, ,
\end{equation}
where we used $d\zeta^k\wedge d\zeta^j \wedge d\zeta^i \equiv \epsilon^{kji} d^3 \zeta$.

\begin{figure}[h]
\centering 
  \includegraphics[scale=.5]{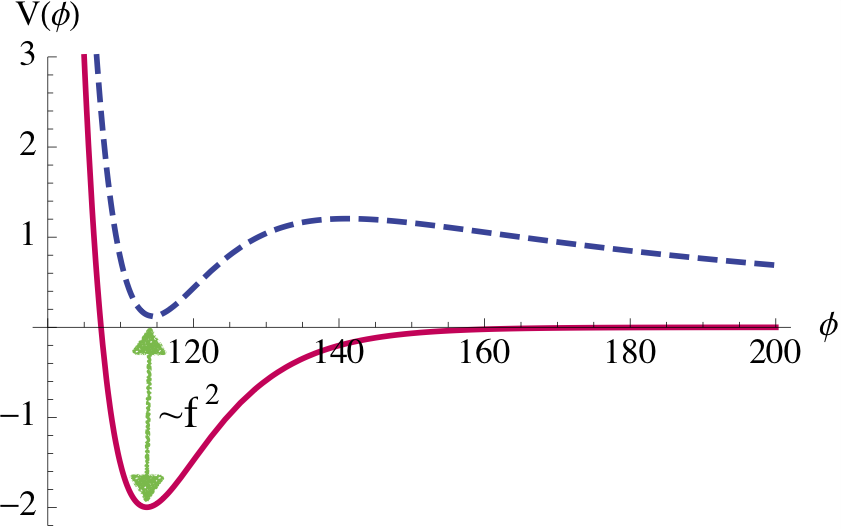}
\caption{ 
{\it The uplift due to the nilpotent goldstino from anti-de Sitter to de Sitter in the KKLT model. This system can decay back to a supersymmetric anti-de Sitter phase if the supersymmetry breaking is supported by a four-form flux. Depending on the membrane tension the decay can be rapid or slow.} \label{KKLT-JUMPS}} 
\end{figure}

Let us assume to have a spherical membrane, such that there is a region outside and a region inside the membrane. 
Integrating the three-form equation we find that the value of ${\rm n} = \langle\star dC_3\rangle$ is different on the sides of the membrane, thus inducing a jump on the flux 
\begin{equation}
\label{Dn}
\Delta {\rm n} \equiv {\rm n}_{\rm out} - {\rm n}_{\rm in} =- \frac{\mu}{8 \pi} \,. 
\end{equation}
If ${\rm n}_{\rm out}=f$, where $f$ is a non-zero real constant and  $\mu=-8\pi f$, 
then ${\rm n}_{\rm in}=0$, which means that the value of ${\rm n} =\langle \partial_\mu C^\mu \rangle $ goes from non-vanishing to vanishing, and supersymmetry gets restored. 
When supersymmetry is restored the full goldstino multiplet has to vanish. 
Indeed, the simplest way to see this is by noticing that the transformation of the goldstino reads 
\begin{equation}
\delta G = f P_L \epsilon + P_L \partial \!\!\!/ \left(\frac{G^2}{2f}\right) \epsilon + \dots .
\end{equation}
Therefore, when $f$ vanishes it enforces the condition 
\begin{equation}
f \to 0 \quad \Rightarrow \quad  G \equiv 0 \,, 
\end{equation}
and the goldstino evaporates while supersymmetry gets restored. 
Note that while the two backgrounds/vacua are disconnected, there can be a bubble of true vacuum growing within the meta-stable state as depicted in the figure below.  
\begin{center}
	\begin{tikzpicture}[node distance=0cm]
	\tikzstyle{blobIN} = [circle, rounded corners, minimum width=3.5cm, minimum height=1cm, text centered, yshift=-0.35cm, 
	label=above:$\langle \star dC_3 \rangle \!\ne\! 0 \ \to \ {\rm NL\!-\!SUSY}$, draw=black, fill=gray!01]
		\tikzstyle{blobOUT} = [rectangle, rounded corners, minimum width=6cm, minimum height=5cm, draw=black, fill=green!09]
		\node (meta) [blobOUT] {}; 	
		\node (true) [blobIN] {$\star dC_3 \!=\! 0 \ \to \ G\!\equiv\!0$}; 
	\end{tikzpicture}   
\end{center}
This has been analyzed in a fully supersymmetric setup in \cite{Farakos:2020wfc}, using superspace methods. In the same work, it has also been shown that four-dimensional de Sitter vacua of $\mathcal{N}=1$ supergravity supported only by the nilpotent goldstino could decay if the supersymmetry breaking is supported by a flux and if the three-form is coupled to a membrane, see figure \ref{KKLT-JUMPS}.

\section{Further constraints from supergravity EFTs} 
\label{sec:newconj}

In this section, we present known but also new swampland conjectures which most clearly draw inspiration and motivation from supergravity. In particular, we review the gravitino conjecture originally put forward in \cite{Cribiori:2021gbf,Castellano:2021yye}, we speculate on the possible existence of a new conjecture related to the Yukawa couplings and we discuss the consequence of applying the Festina Lente bound, reviewed in section \ref{sec:WGCandFL}, to D-term inflation.\footnote{Within this logic one could also study what happens if we impose the A-TCC condition \cite{Andriot:2022brg} on four-dimensional supergravity theories and search for new constraints.} The material contained in sections \ref{sec:yuk} and \ref{sec:FLvsDterminf} is new and it has not been published elsewhere.

\subsection{Gravitino conjecture}

The gravitino is perhaps the central building block of any supergravity theory and a considerable amount of information can be deduce just by looking at its interactions. Indeed, we exploited this fact already in section \ref{sec:WGCvs}, where the gravitino played a central role in applying the WGC to supergravity models.

A swampland conjecture exclusively centred around the gravitino has been proposed in \cite{Cribiori:2021gbf,Castellano:2021yye}. It holds that the limit of vanishing gravitino mass,
\begin{equation}
m_{3/2} \to 0,
\end{equation}
is accompanied by an infinite tower of states becoming light and invalidating the effective description. 
In simple enough cases, one expects the typical mass $m$ of these states to scale with a power of the gravitino mass as
\begin{equation}
\label{mm32n}
m\sim \left(\frac{m_{3/2}}{M_P}\right)^n M_P,
\end{equation}
where $n$ is a model dependent positive parameter of order one.

Notice that for supersymmetric anti-de Sitter vacua this statement is equivalent to the anti-de Sitter distance conjecture of \cite{Lust:2019zwm}. However, the gravitino conjecture is proposed to be valid on any background, not only on supersymmetric anti-de Sitter phases, and perhaps its most relevant application for phenomenology is to non-supersymmetric vacua with an almost vanishing cosmological constant, as the one measured today. Indeed, in these setups the general structure of supergravity tells us that the gravitino mass is related to the supersymmetry breaking scale as
\begin{equation}
M^2_{SUSY} \simeq m_{3/2} M_P,
\end{equation}
and thus one may hope to learn something about the scale at which supersymmetry is broken in our universe by investigating the domain of validity of the effective theory describing it. In \cite{Anchordoqui:2023oqm}, this aspect has been analyzed further in the context of the Dark Dimension scenario \cite{Montero:2022prj}.

The gravitino conjecture can be tested directly in various supergravity effective models. A natural ultraviolet cutoff in supergravity is the Kaluza-Klein scale and thus one can expect the states predicted by the gravitino conjecture to be Kaluza-Klein states.\footnote{More in general, a well-defined notion of ultraviolet cutoff for the gravitation interactions is the so called species scale \cite{Dvali:2007hz,Dvali:2007wp,Dvali:2009ks,Dvali:2010vm,Dvali:2012uq}. Its interplay with the gravitino conjecture has been studied already in the original paper \cite{Cribiori:2021gbf}.} For an isotropic compact manifold with volume $\mathcal{V}$, one has $m_{KK}\sim \mathcal{V}^{-2/3}M_P$. On the other hand, for
Calabi-Yau and orbifold compactifications the volume $\mathcal{V}$ enters the K\"ahler potential as
\begin{equation}
K = -\alpha\log \mathcal{V} + K',
\end{equation}
for some model-dependent parameter $\alpha$. 
Here, $\mathcal{V}$ is a function of the K\"ahler moduli, while the remaining part $K'$ depends on the complex structure moduli and the axio-dilaton. Assuming that at the critical point of the scalar potential the superpotential scales as $W\sim \mathcal{V}^{\frac{\beta}{2}}$, one derives 
\begin{equation}
m_{3/2}\sim \mathcal{V}^{\frac{\beta-\alpha}{2}}. 
\end{equation}
By comparing with \eqref{mm32n}, one identifies the parameter of the conjecture with
\begin{equation}
n=\frac{4}{3(\alpha-\beta)},
\end{equation}
and it can then be calculated in concrete examples \cite{Cribiori:2021gbf,Castellano:2021yye}. For instance, in heterotic compactifications one finds $n=4/3$, while for GKP orientifolds $n=2/3$, and for Scherk-Schwarz compactifications $n=4$. Models with non-perturbative superpotential and in which the gravitino conjecture is dynamically realized have been pointed out in \cite{Cribiori:2022sxf}. Additional support to the conjecture stems from $\mathcal{N}=2$ supergravity, as discussed in \cite{Cribiori:2021gbf}. Indeed, in the STU model arising e.g.~from heterotic compactifications on $K3\times T^2$, one finds a one to one relation between the gravitino mass and the gauge coupling, $m_{3/2}\simeq g_{3/2}$, which allows to connect the gravitino conjecture to the WGC and to the absence of global symmetries in quantum gravity. 
Further evidence for the gravitino conjecture in string theory has been provided in \cite{Coudarchet:2021qwc}. These constructions have the peculiar feature that the gravitino mass can be decoupled from the cosmological constant, while avoiding open strings tachyonic instabilities. It is then shown that the limit of vanishing gravitino mass is inconsistent, for it clashes with the boundary conditions of the gravitinos imposed by the simultaneous presence of orientifold and anti-orientifold planes.

\subsection{Constraints on Yukawa couplings?} 
\label{sec:yuk}

In this section, we focus on Yukawa couplings $Y$ and we would like to analyze if and how supergravity can hint at swampland restrictions on them. A related discussion can be also found in \cite{Castellano:2022bvr}. 
As we will see, we can deduce a condition of the form 
\begin{equation}
\label{Yuka-1}
\Lambda_{UV} \lesssim Y^{\alpha} M_P \,, 
\end{equation}
for some parameter $\alpha$. At the present stage, we do not know if such a condition is an accident that happens to hold only in few instances or if it has a deeper origin within string theory and therefore can constitute an actual swampland conjecture. In case the latter option turns out to be correct, our discussion can be seen as yet another confirmation of the fact that supergravity is in many circumstances closer to quantum gravity than an ordinary effective theory.
In addition, here we cannot tell for certain what is the order of magnitude of the parameter $\alpha$.

We will give two examples where supergravity supplemented by a given swampland conjecture leads to the condition \eqref{Yuka-1}. 
Note that, due to \eqref{Yuka-1}, one can further speculate that the limit
\begin{equation}
\label{Yuka-2}
Y \to 0  
\end{equation}
corresponds to travelling to infinite distance in the moduli space. Indeed, the two conditions \eqref{Yuka-1} and \eqref{Yuka-2} can be  related to each other since as we go to infinite distance we expect the cut-off to be lowered and towers of states to become massless. Then, the parameter $\alpha$ would be identified by the nature of states becoming light. 
Notice that the above constraints are not really the bounds studied in \cite{Palti:2020tsy}, but do have a similar gist to it, therefore one can speculate that there might be an underlying connection.

First, let us consider the simple no-scale model 
\begin{equation}
K = -3 \log(T + \bar T) \quad , \quad W = W_0 \,. 
\end{equation}
The idea is to relate the Yukawa coupling of the scalar $T$ to the gravitino mass. Then, enforcing the gravitino conjecture we can get a constraint on the Yukawa coupling itself. In other words, we will see that the gravitino conjecture automatically becomes a conjecture on the Yukawa coupling. 
The gravitino mass is given by
\begin{equation}
m_{3/2} = \frac{W_0}{(T + \bar T)^{3/2}} \,. 
\end{equation}
Due to the no-scale structure, the scalar potential is identically vanishing. The kinetic term of $T$ is
\begin{equation}
e^{-1} {\cal L}_{kin} = - \frac{3}{(T + \bar T)^2} \partial_\mu T \partial^\mu \bar T  
= - \frac12 \partial_\mu \phi \partial^\mu \phi + \dots
\end{equation}
where we have introduced the canonically normalized real scalar $\phi$ defined as
\begin{equation}
{\rm Re }T =  e^{\sqrt{\frac{2}{3}} \phi} \,. 
\end{equation}
In terms of $\phi$, the gravitino mass reads 
\begin{equation}
m_{3/2} = 2^{-3/2} W_0 e^{ - \sqrt \frac32 \phi} \,. 
\end{equation}
This produces a Yukawa coupling of the form 
\begin{equation}
Y M_P = - \sqrt \frac32 m_{3/2}.
\end{equation}
By enforcing the gravitino conjecture \eqref{mm32n}, we expect a tower of states to become light in the limit $Y\to 0$, invalidating the effective description. Notice that the relation between the Yukawa coupling and the gravitino mass is a consequence of the supersymmetric structure of the lagrangian and might not hold outside the realm of supergravity.

The second instance where the Yukawa coupling takes a suggestive form is when considering gauged $\mathcal{N}=1$ supergravity. 
Once again, let us discuss the simplest case where this effect can happen. 
We consider a flat K\"ahler potential 
\begin{equation}
K = z \overline z,
\end{equation}
with a typical U$(1)$ gauging with Killing vector $k=iz$ and constant gauge kinetic function $f = 1 /g^2$. 
Then, an inspection of the component Lagrangian reveals that supersymmetry requires a term of the form 
\begin{equation}
g z \, \overline \chi P_L  \lambda \,, 
\end{equation}
where $\chi$ is the fermionic superpartner of $z$ and $\lambda$ the gaugino.
This is clearly a Yukawa coupling between the scalar and the fermions.  Enforcing  the magnetic WGC, we eventually have that
\begin{equation}
Y M_P = g M_P \sim \Lambda_{WGC} \,,
\end{equation}
which is along the same lines of \eqref{Yuka-1}.

We conclude that supergravity readily gives indications that Yukawa couplings are often restricted by 
swampland-like constraints that have the form \eqref{Yuka-1} and therefore also imply \eqref{Yuka-2}. 
This may be an actual independent swampland constraint on its own or simply an accidental restriction due to supersymmetry, however our experience until now hints to the former possibility. 
Nevertheless, we do not have enough evidence to further support the existence of such a conjecture for the time being, nor do we know its exact form or the expected values of the parameter $\alpha$.

\subsection{Festina Lente versus D-term inflation} 
\label{sec:FLvsDterminf}

We have seen in the discussion on $\mathcal{N}=1$ and $\mathcal{N}=2$ de Sitter vacua in section \ref{sec:WGCvs} that the magnetic WGC or the Festina Lente bound can equally well raise questions about the validity of the two-derivative approximation within the effective theory. 
One further step would be to investigate the extension of such restrictions to dynamical systems, as for example supergravity models of inflation \cite{Linde:1990flp,Lyth:1998xn}. 
The bibliography on supergravity cosmology is vast and it is not our purpose to review it here, but it is fair to say that in principle there are two concrete methods allowing the construction of single-field inflationary models within four-dimensional $\mathcal{N}=1$ supergravity. 
These are related to the basic source of the energy density driving inflation: it can be either a D-term or an F-term. 
The former is for situations where inflation is essentially controlled by the potential due to a gauging, whereas the latter is for when the inflationary potential is controlled by the superpotential. 
The most general constructions for single-field inflationary models with vector multiplets, i.e.~D-term models, can be found in \cite{Ferrara:2013rsa}. They are based on the supergravity inflationary model that first appeared in \cite{Farakos:2013cqa}, 
whereas the single-field inflationary models with superpotentials, i.e.~F-term models, can be found in \cite{Kallosh:2010xz}. 
The latter are based on the ideas that first appeared in \cite{Kawasaki:2000yn}.\footnote{However, the first successful models that utilize the idea of chaotic inflation in supergravity where constructed quite early in the literature \cite{Goncharov:1983mw}.} 
Efforts to construct inflationary models in $\mathcal{N}=2$ supergravity with similar logic were made in \cite{Ceresole:2014vpa}. 
Here, we will show that, modulo some caveats that we highlight, the Festina Lente bound can pose strong constraints on the D-term models, whereas we can only speculate for similar restrictions for the F-term models.\footnote{There are of course already interesting alternatives, see e.g. \cite{Dalianis:2014sqa,DallAgata:2018ybl,DallAgata:2019yrr} for supergravity embedding, that should be studied in more detail in case the simple way to realize inflation/dark energy within supergravity fails. Constraints imposed on inflation by the swampland distance conjecture have been studied e.g.~in \cite{Scalisi:2018eaz}, but they turn out to be milder than those that we derive below by imposing Festina Lente.}

Let us recap the basic requirements for single-field inflation in supergravity and its challenges, which become even worse when one has to find an embedding into string theory. This is possibly reflected in the difficulty of reconciling such models with the constraints coming from the swampland program. 
First of all, taking into account that we introduce various multiplets to be able to build the desired scalar potential, during inflation it is crucial to have a so-called strong stabilization of all scalars except the inflaton. 
This means that all extra scalars should get masses safely above the Hubble scale, $H = \sqrt{V/3}$ in Planck units. 
Subsequently one focuses on the potential for the single scalar, say $\phi$, that plays the role of the inflaton and studies the slow-roll conditions. This field has typically a lagrangian 
\begin{equation}
e^{-1} {\cal L} = \frac12 R - \frac12 w(\phi)\, \partial_\mu \phi \partial^\mu\phi - V(\phi) \,
\end{equation}
and for our analysis here it is enough to look at the so-called $\epsilon$ slow-roll condition 
\begin{equation}
\epsilon(\phi) \Big{|}_{\text{inflation}} \equiv \frac{1}{2 w(\phi)} \left(\frac{V'}{V}\right)^2 \Big{|}_{\text{inflation}} \ll 1 \,,
\end{equation}
while inflation conventionally ends when $\epsilon (\phi_{\text{end}}) \simeq 1$. 
This condition essentially means the potential is not steep in the region where inflation takes place. 
There is also one further slow-roll condition that restricts $V''$, but we do not need to invoke it here; it can be found for example in \cite{Dalianis:2014aya} for general $V(\phi)$ and $w(\phi)$. 
Even if we do have such a potential, there are various other issues that have to be addressed, the zeroth order test being to match with observations \cite{Planck:2018jri}. 
Further theoretical issues that need to be addressed are the initial conditions problem \cite{Goldwirth:1991rj,Ijjas:2013vea,Dalianis:2015fpa,Linde:2017pwt}, or the problem of higher order corrections \cite{Farakos:2013cqa,Ferrara:2013kca}. 
However, we will not look into these problems here.

\subsubsection{General restrictions from Festina Lente}

A proposal that allows to build versatile single-field models of D-term inflation was presented in \cite{Farakos:2013cqa,Ferrara:2013rsa} 
and it is technically based on models with massive vector multiplets \cite{VanProeyen:1979ks,Cecotti:1987qr,Cecotti:1987qe}. 
In the setup we have two multiplets, a chiral multiplet $z$ and an abelian gauge multiplet, while the superpotential vanishes. 
The isometry we want to gauge is a shift 
\begin{equation}
z \to z + i \alpha \,, 
\end{equation}
which means that the gauge field is essentially massive and therefore it will absorb one of the two real scalars residing in $z$, leaving behind only one real scalar that becomes the inflaton. That the gauge field is massive can be seen from the covariant derivative of the scalar \eqref{covderz} which, for the shift symmetry described here, reads $D_\mu z = \partial_\mu z-iA_\mu$, leading to a mass term for $A_\mu$.
Indeed, from the definition of the Killing vectors \eqref{defkillv} we see that $k=i$ which means that the real moment map is such
\begin{equation}
\partial_{\bar z} {\cal P} = - g_{z \bar z}  
\end{equation}
giving in turn
\begin{equation}
{\cal P} = - K_z + \xi \,. 
\end{equation}
From this, one deduces that the K\"ahler potential has the form $K=K(z + \bar z)$ and the real constant $\xi$ is the Fayet--Iliopoulos term. 
Since the gauge vector will eventually absorb the imaginary part of $z$, we can readily go to the unitary gauge and set $z$ to a real scalar $Z$, such that we have $D_\mu z = (1/2) \partial_\mu Z - i A_\mu$. 
Furthermore, we choose the gauge kinetic function to be just a constant, $f=1/g^2$. 
From the property that the K\"ahler potential depends only on the real part $z + \bar z = Z$, we have $g_{z \bar z} = J''(Z)>0$ and $K_{z} = K_{\bar z} = J'(Z)$, where we introduced $J(Z) = K|_{z+\overline z = Z}$. 
Taking all of this into account, we rewrite the moment map as
\begin{equation} 
{\cal P}(Z) = \xi - J'(Z) \,,
\end{equation}
and the bosonic sector of the Lagrangian, in the abelian unitary gauge, reads 
\begin{equation}
e^{-1} {\cal L} = \frac12 R 
- \frac14 J''(Z) \partial_\mu Z \partial^\mu Z 
- J''(Z) A_\mu A^\mu
- \frac{1}{4 g^2} F_{\mu\nu}F^{\mu\nu} - \frac12 g^2 {\cal P}^2(Z) \,. 
\end{equation}
Most importantly, we also see that the gravitino has charge $q_{3/2}$ under $A_\mu$ and mass given by 
\begin{equation}
q_{3/2} = {\cal P}(Z) , \qquad m_{3/2} \equiv  0 \,, 
\end{equation}
implying that during inflation, which can only take place for ${\cal P}(Z) \ne 0$, the gravitino is in fact charged and massless.\footnote{Here, we are taking the charge of the gravitino to be a function of the scalar fields, ${\cal P}(Z)$. In general, the physical coupling is the product of the charge with the gauge coupling, $gq$, 
which is in principle field-dependent. Similarly, here we have $g q_{3/2}= g {\cal P}(Z)$, namely  a field-dependent coupling 
between the gravitino and the vector (which is in any case massive in this model).}

At this point, one could naively apply the Festina Lente bound right away, but that would not be correct. This is because the vector $A_\mu$ is massive and so the bound does not apply directly. However, according to \cite{Montero:2021otb} the bound can apply when the mass of the vector is safely below the Hubble scale, that is when 
\begin{equation}
\frac{m_A}{H} < 1.
\end{equation}
Hence, when the vector is that light, one can apply the Festina Lente bound and conclude that because the gravitino is charged but massless this class of inflationary models belongs to the swampland, or at least such a region of the scalar potential cannot be trusted. 
Let us therefore check under which conditions the vector becomes parametrically light. 
In this model, the mass of the vector is given by 
\begin{equation}
m^2_A = 2 g^2 J''(Z) \,. 
\end{equation}
It is then instructive to rewrite the slow-roll $\epsilon$-parameter as
\begin{equation}
\epsilon(Z) \equiv \frac{1}{2 w(Z)} \left(\frac{V'(Z)}{V(Z)}\right)^2 
= \frac{2 g^2 J''}{\frac12 g^2 {\cal P}^2} = \frac{m_A^2}{3 H^2} \,. 
\end{equation}
From this relation, one can see directly that the slow-roll regime $\epsilon \ll 1$ corresponds precisely to a parametrically light vector, namely
\begin{equation}
\epsilon \ll1 \qquad \Longleftrightarrow\qquad \frac{m_A^2}{3 H^2} \ll 1 ,
\end{equation}
and thus, within such a regime the Festina Lente bound applies and relegates the inflating region into the swampland. 
Instead of invoking a massless charged gravitino, one can arrive at an analogous result by comparing the vacuum energy of the system to the bound imposed by Festina Lente.

One can also ask if a similar conclusion can be reached by using the magnetic WGC. 
In the limit where we are deep in the regime of slow-roll then the system effectively behaves like a model with a pure Fayet-Iliopoulos term and a massless gravitino, and so one could already expect that the results found in \cite{Cribiori:2020wch} apply and set the model in the swampland.

Let us stress that this result has a series of caveats that should be studied in detail. 
First of all, one should test carefully that the Festina Lente bound is to be trusted on dynamical backgrounds 
as for example an inflating universe. 
Second, since the original bound is for massless vectors, one should verify that the Festina Lente bound can be concretely trusted for (parametrically) light vectors whose mass is below the Hubble scale. 
Clearly, for an observer living within a causally connected region within the de Sitter Hubble horizon, a vector with mass smaller than the Hubble scale is effectively massless and it acts as a long-range force.
Thirdly, in a realistic setup one should unavoidably consider including a superpotential $W$ and it is therefore possible that, maybe with a 
judicious choice of $W$, inflation can be safe. Either way, the clean simple model discussed above seems to be in conflict with the Festina Lente bound.

\subsubsection{Examples: Chaotic and Starobinsky inflation}

Before, we required $\epsilon \ll 1$ since we were interested in a situation where inflation is safe and can happen for as many e-folds as one wants. Now, we turn to regimes in which slow-roll has marginal validity and takes place for a very restricted amount of e-folds. 
Indeed, one could ask if it is possible to satisfy slow-roll while staying outside of the swampland, that is while keeping the vector's mass above $H$, even for a restricted number of e-folds. 
This requirement implies the inflationary trajectory is bounded by two conditions. 
First, we know that inflation ends near $\epsilon \simeq 1$, 
which means that we have 
\begin{equation}
\epsilon |_{\text{end}} = 1 \quad \to \quad m_A^2(Z_{\text{end}} ) = 3 H^2 (Z_{\text{end}} ) \,.
\end{equation}
As we discussed, the Festina Lente bound (assuming always it applies to massive vectors) places the inflationary phase in the swampland when $m_A < H$, so we can be generous about the validity of the effective description and say that $Z_\text{FL}$ corresponds to an upper bound value of $m_A$ given, say, by 
\begin{equation}
m_A (Z_\text{FL}) \simeq \frac{1}{6} \times H(Z_\text{FL}) \,. 
\end{equation} 
Then, we conclude that inflation at best can take place within the region 
\begin{equation}
Z_{\text{end}} < Z_{\text{inflation}} < Z_{\text{FL}} \, 
\end{equation}
and the question is if there is enough amount of distance that can be travelled by the inflaton within such bound 
and how many e-folds of inflation we can have. 
We will investigate this question below by studying two inflationary models: the typical quadratic chaotic inflation and the Starobinsky model. 
Notably these two models are limiting cases of the so-called $\alpha$-attractor models \cite{Ferrara:2013rsa,Kallosh:2013yoa}, 
therefore one might expect $\alpha$-attractors to have similar bounds as well.

\paragraph{Quadratic chaotic inflation.}
We consider first the quadratic chaotic inflation model, which is however currently in tension with observations from the Planck satellite. It requires 
\begin{equation}
J(Z) = Z^2, \qquad \xi = 0 \,. 
\end{equation}
The scalar and the vector fields have a canonical kinetic terms, while the scalar potential and the vector mass take the form 
\begin{equation}
V = 2 g^2 Z^2, \qquad m_A^2 = 4 g^2 \,, 
\end{equation}
giving in turn
\begin{equation}
m_A = \frac{\sqrt 6}{Z}  H, \qquad \epsilon = \frac{2}{Z^2} \,. 
\end{equation}
The gravitino is massless and charged with $q_{3/2} = -2 Z$. 
We find the boundaries of the inflating region to be 
\begin{equation}
Z_\text{end} = \sqrt 2 , \qquad  Z_{\text{FL}} \simeq 6  \sqrt 6 \,. 
\end{equation}
The number of e-folds then reads 
\begin{equation}
N_\text{e-folds} = \int_{Z_\text{end}}^{Z_{\text{FL}}} \frac{1}{\sqrt{2 \epsilon(Z)}} dZ 
= \frac12 \int_{\sqrt 2}^{6  \sqrt 6} Z dZ \simeq 54 \,. 
\end{equation}
At best, we see that inflation will only produce a very limited amount of e-folds. 
If we assume that $Z_{\text{FL}}$ is instead determined by a stronger constraint, say $m_A (Z_\text{FL}) = \frac{1}{3} \times H(Z_\text{FL})$, then we get a much smaller amount of e-folds, say $N_\text{e-folds} = 13$.

\begin{figure}
\centering 
  \includegraphics[scale=.4]{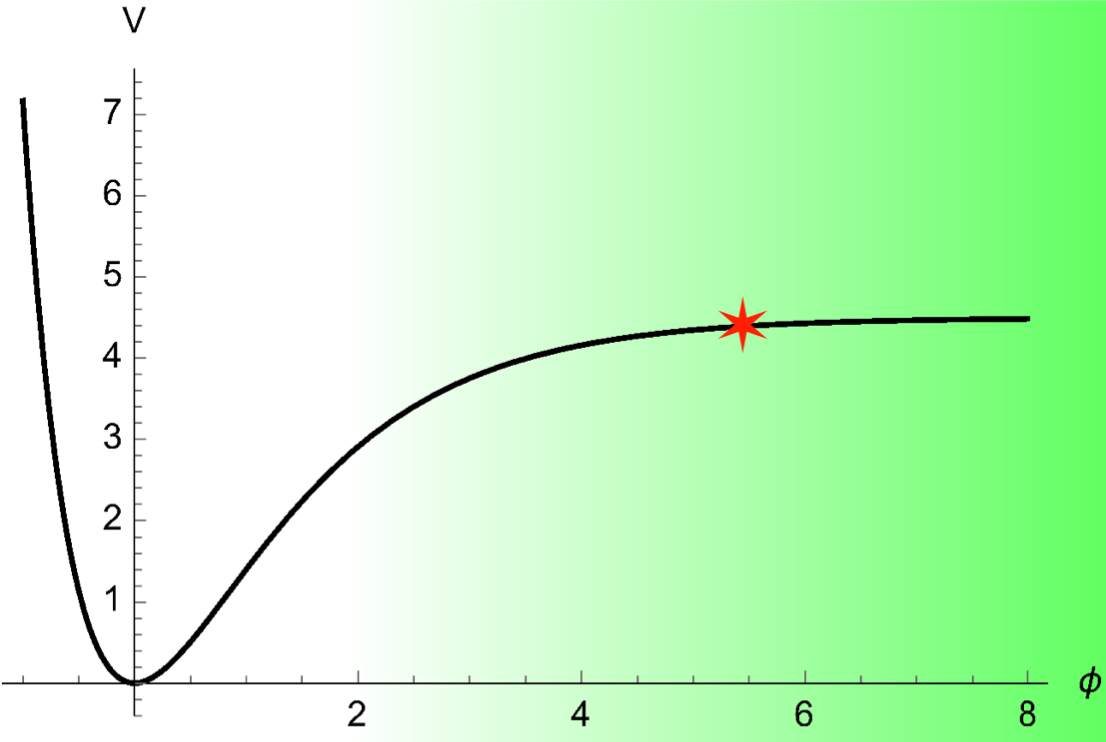}
\caption{ 
{\it Plot of a plateau-type inflationary potential. The red star corresponds to the minimal position where inflation can start in order to eventually give a reasonable number of e-folds, whereas the green regions correspond to regimes that are more and more into the swampland due to the Festina Lente bound. 
One can interpret this effect as the way that supergravity relates the Festina Lente bound to the Distance Conjecture.} \label{Plot-Staro}} 
\end{figure}

\paragraph{Starobinsky model of inflation.}
Next, we check a model of plateau inflation which is currently favored by Planck data. 
For the Starobinsky model we have 
\begin{equation}
J = - 3 \left( \log(-Z) + Z\right) , \qquad \xi = 0 \,. 
\end{equation}
This leads to a bosonic sector of the form
\begin{equation}
e^{-1} {\cal L} = \frac12 R 
- \frac{3}{4 Z^2} \partial_\mu Z\partial^\mu Z  
- \frac{3}{Z^2} A_\mu A^\mu
- \frac{1}{4 g^2} F_{\mu\nu}F^{\mu\nu} 
- \frac{9}{2} g^2 \left( 1 + \frac{1}{Z} \right)^2 \,.  
\end{equation}
Note that the way inflation works here is similar in logic as the method discussed in \cite{Stewart:1994ts}. 
To go to a canonically normalized scalar we set 
$Z = - e^{\sqrt{\frac{2}{3}} \, \phi}$. The scalar potential and the vector mass become 
\begin{equation}
V = \frac{9}{2} g^2 \left( 1 - e^{-\sqrt{\frac{2}{3}} \, \phi }\right)^2 , \qquad 
m_A^2 = 6 g^2 e^{- 2 \sqrt{\frac{2}{3}} \, \phi} \,. 
\end{equation}
From our previous discussion we find that now 
\begin{equation}
\phi_\text{end} \simeq 0.94, \qquad  \phi_{\text{FL}} \simeq 3.15 \,.  
\end{equation}
The number of e-folds then reads 
\begin{equation}
N_\text{e-folds} = \int_{\phi_\text{end}}^{\phi_{\text{FL}}} \frac{1}{\sqrt{2 \epsilon(\phi)}} d\phi \simeq 6.85  \,. 
\end{equation}
We see that the bound is stronger on models of plateau inflation and so one would have to assume a weaker 
onset of the Festina Lente bound than $m_A (Z_\text{FL}) \simeq \frac{1}{6} \times H(Z_\text{FL})$ to get a larger number of e-folds. 
Indeed, 
if we assume $m_A (Z_\text{FL}) \simeq \frac{1}{40} \times H(Z_\text{FL})$, 
then we find $Z_\text{FL} \simeq 5.382$, 
which gives $N_\text{e-folds} \simeq 56.4$, marginally compatible with the Planck data.

Actually, it seems that if we extrapolate the results to small-field inflation, the bound from Festina Lente will be severe. 
This happens because a very flat region is required for inflation to take place which automatically means that the vector will be extremely light thus Festina Lente will apply and set all the inflating regime in the swampland.

\section*{Acknowledgements} 

We thank I.~Bandos, G.~Dall'Agata, M.~Emelin, A.~Kehagias, N.~Liatsos, D.~L\"ust, L.~Martucci, M.~Morittu, M.~Scalisi, D.~Sorokin, and G.~Tringas for collaboration and discussions that led to the works reviewed here. We thank T.~Van Riet for encouraging us in writing this review. The work of N.C.~is supported by the Alexander-Von-Humboldt foundation. The work of F.F.~is supported by the MIUR-PRIN contract 2017CC72MK003.


\begin{thebibliography}{99}
\bibitem{Vafa:2005ui}
C.~Vafa,
``The String landscape and the swampland,''
[arXiv:hep-th/0509212 [hep-th]].


\bibitem{Palti:2019pca}
E.~Palti,
``The Swampland: Introduction and Review,''
Fortsch. Phys. \textbf{67} (2019) no.6, 1900037
[arXiv:1903.06239 [hep-th]].

\bibitem{Agmon:2022thq}
N.~B.~Agmon, A.~Bedroya, M.~J.~Kang and C.~Vafa,
``Lectures on the string landscape and the Swampland,''
[arXiv:2212.06187 [hep-th]].

\bibitem{Freedman:1976xh}
D.~Z.~Freedman, P.~van Nieuwenhuizen and S.~Ferrara,
``Progress Toward a Theory of Supergravity,''
Phys. Rev. D \textbf{13} (1976), 3214-3218

\bibitem{Deser:1976eh}
S.~Deser and B.~Zumino,
``Consistent Supergravity,''
Phys. Lett. B \textbf{62} (1976), 335

\bibitem{Wess:1992cp}
J.~Wess and J.~Bagger,
``Supersymmetry and supergravity,''
Princeton University Press, 1992,
ISBN 978-0-691-02530-8

\bibitem{Freedman:2012zz}
D.~Z.~Freedman and A.~Van Proeyen,
``Supergravity,''
Cambridge Univ. Press, 2012,
ISBN 978-1-139-36806-3, 978-0-521-19401-3


\bibitem{DallAgata:2021uvl}
G.~Dall\textquoteright{}Agata and M.~Zagermann,
``Supergravity: From First Principles to Modern Applications,''
Lect. Notes Phys. \textbf{991} (2021), 1-263
2021,
ISBN 978-3-662-63978-8, 978-3-662-63980-1

\bibitem{Cecotti:1985mf}
S.~Cecotti, S.~Ferrara, L.~Girardello, M.~Porrati and A.~Pasquinucci,
``Matter Coupling in Higher Derivative Supergravity,''
Phys. Rev. D \textbf{33} (1986), 2504


\bibitem{Cecotti:1987mr}
S.~Cecotti, S.~Ferrara, L.~Girardello, A.~Pasquinucci and M.~Porrati,
``Matter Coupled Supergravity With {Gauss-Bonnet} Invariants: Component Lagrangian and Supersymmetry Breaking,''
Int. J. Mod. Phys. A \textbf{3} (1988), 1675-1733

\bibitem{Buchbinder:1988tj}
I.~L.~Buchbinder and S.~M.~Kuzenko,
``QUANTIZATION OF THE CLASSICALLY EQUIVALENT THEORIES IN THE SUPERSPACE OF SIMPLE SUPERGRAVITY AND QUANTUM EQUIVALENCE,''
Nucl. Phys. B \textbf{308} (1988), 162-190

\bibitem{Buchbinder:1998qv}
I.~L.~Buchbinder and S.~M.~Kuzenko,
``Ideas and methods of supersymmetry and supergravity: Or a walk through superspace,''

\bibitem{Andrianopoli:1996cm}
L.~Andrianopoli, M.~Bertolini, A.~Ceresole, R.~D'Auria, S.~Ferrara, P.~Fre and T.~Magri,
``N=2 supergravity and N=2 superYang-Mills theory on general scalar manifolds: Symplectic covariance, gaugings and the momentum map,''
J. Geom. Phys. \textbf{23} (1997), 111-189
[arXiv:hep-th/9605032 [hep-th]].

\bibitem{Ceresole:1995ca}
A.~Ceresole, R.~D'Auria and S.~Ferrara,
``The Symplectic structure of N=2 supergravity and its central extension,''
Nucl. Phys. B Proc. Suppl. \textbf{46} (1996), 67-74
[arXiv:hep-th/9509160 [hep-th]].

\bibitem{Craps:1997gp}
B.~Craps, F.~Roose, W.~Troost and A.~Van Proeyen,
``What is special Kahler geometry?,''
Nucl. Phys. B \textbf{503} (1997), 565-613
[arXiv:hep-th/9703082 [hep-th]].

\bibitem{Arkani-Hamed:2006emk}
N.~Arkani-Hamed, L.~Motl, A.~Nicolis and C.~Vafa,
``The String landscape, black holes and gravity as the weakest force,''
JHEP \textbf{06} (2007), 060
[arXiv:hep-th/0601001 [hep-th]].

\bibitem{Dvali:2007hz}
G.~Dvali,
``Black Holes and Large N Species Solution to the Hierarchy Problem,''
Fortsch. Phys. \textbf{58} (2010), 528-536
[arXiv:0706.2050 [hep-th]].

\bibitem{Dvali:2007wp}
G.~Dvali and M.~Redi,
``Black Hole Bound on the Number of Species and Quantum Gravity at LHC,''
Phys. Rev. D \textbf{77} (2008), 045027
[arXiv:0710.4344 [hep-th]].

\bibitem{Dvali:2009ks}
G.~Dvali and D.~Lust,
``Evaporation of Microscopic Black Holes in String Theory and the Bound on Species,''
Fortsch. Phys. \textbf{58} (2010), 505-527
[arXiv:0912.3167 [hep-th]].

\bibitem{Dvali:2010vm}
G.~Dvali and C.~Gomez,
``Species and Strings,''
[arXiv:1004.3744 [hep-th]].

\bibitem{Dvali:2012uq}
G.~Dvali, C.~Gomez and D.~Lust,
``Black Hole Quantum Mechanics in the Presence of Species,''
Fortsch. Phys. \textbf{61} (2013), 768-778
[arXiv:1206.2365 [hep-th]].

\bibitem{Huang:2006hc}
Q.~G.~Huang, M.~Li and W.~Song,
``Weak gravity conjecture in the asymptotical dS and AdS background,''
JHEP \textbf{10} (2006), 059
[arXiv:hep-th/0603127 [hep-th]].

\bibitem{Antoniadis:2020xso}
I.~Antoniadis and K.~Benakli,
``Weak Gravity Conjecture in de Sitter Space-Time,''
Fortsch. Phys. \textbf{68} (2020) no.9, 2000054
[arXiv:2006.12512 [hep-th]].


\bibitem{Cribiori:2022trc}
N.~Cribiori and G.~Dall'Agata,
``Weak gravity versus scale separation,''
JHEP \textbf{06} (2022), 006
[arXiv:2203.05559 [hep-th]].

\bibitem{Lust:2004ig}
D.~Lust and D.~Tsimpis,
``Supersymmetric AdS(4) compactifications of IIA supergravity,''
JHEP \textbf{02} (2005), 027
[arXiv:hep-th/0412250 [hep-th]].

\bibitem{Tsimpis:2012tu}
D.~Tsimpis,
``Supersymmetric AdS vacua and separation of scales,''
JHEP \textbf{08} (2012), 142
[arXiv:1206.5900 [hep-th]].

\bibitem{Gautason:2015tig}
F.~F.~Gautason, M.~Schillo, T.~Van Riet and M.~Williams,
``Remarks on scale separation in flux vacua,''
JHEP \textbf{03} (2016), 061
[arXiv:1512.00457 [hep-th]].

\bibitem{Gautason:2018gln}
F.~F.~Gautason, V.~Van Hemelryck and T.~Van Riet,
``The Tension between 10D Supergravity and dS Uplifts,''
Fortsch. Phys. \textbf{67} (2019) no.1-2, 1800091
[arXiv:1810.08518 [hep-th]].

\bibitem{Lust:2020npd}
D.~L\"ust and D.~Tsimpis,
``AdS$_{2}$ type-IIA solutions and scale separation,''
JHEP \textbf{07} (2020), 060
[arXiv:2004.07582 [hep-th]].

\bibitem{DeLuca:2021mcj}
G.~B.~De Luca and A.~Tomasiello,
``Leaps and bounds towards scale separation,''
JHEP \textbf{12} (2021), 086
[arXiv:2104.12773 [hep-th]].


\bibitem{Collins:2022nux}
T.~C.~Collins, D.~Jafferis, C.~Vafa, K.~Xu and S.~T.~Yau,
``On Upper Bounds in Dimension Gaps of CFT's,''
[arXiv:2201.03660 [hep-th]].

\bibitem{Andriot:2022yyj}
D.~Andriot, L.~Horer and P.~Marconnet,
``Exploring the landscape of (anti-) de Sitter and Minkowski solutions: group manifolds, stability and scale separation,''
JHEP \textbf{08} (2022), 109
[erratum: JHEP \textbf{09} (2022), 184]
[arXiv:2204.05327 [hep-th]].

\bibitem{Lust:2019zwm}
D.~L\"ust, E.~Palti and C.~Vafa,
``AdS and the Swampland,''
Phys. Lett. B \textbf{797} (2019), 134867
[arXiv:1906.05225 [hep-th]].

\bibitem{Blumenhagen:2019vgj}
R.~Blumenhagen, M.~Brinkmann and A.~Makridou,
``Quantum Log-Corrections to Swampland Conjectures,''
JHEP \textbf{02} (2020), 064
[arXiv:1910.10185 [hep-th]].



\bibitem{Buratti:2020kda}
G.~Buratti, J.~Calderon, A.~Mininno and A.~M.~Uranga,
``Discrete Symmetries, Weak Coupling Conjecture and Scale Separation in AdS Vacua,''
JHEP \textbf{06} (2020), 083
[arXiv:2003.09740 [hep-th]].


\bibitem{Emelin:2020buq}
M.~Emelin,
``Effective Theories as Truncated Trans-Series and Scale Separated Compactifications,''
JHEP \textbf{11} (2020), 144
[arXiv:2005.11421 [hep-th]].


\bibitem{Shiu:2022oti}
G.~Shiu, F.~Tonioni, V.~Van Hemelryck and T.~Van Riet,
``AdS scale separation and the distance conjecture,''
[arXiv:2212.06169 [hep-th]].

\bibitem{DeWolfe:2005uu}
O.~DeWolfe, A.~Giryavets, S.~Kachru and W.~Taylor,
``Type IIA moduli stabilization,''
JHEP \textbf{07} (2005), 066
[arXiv:hep-th/0505160 [hep-th]].

\bibitem{Behrndt:2004km}
K.~Behrndt and M.~Cvetic,
``General N = 1 supersymmetric flux vacua of (massive) type IIA string theory,''
Phys. Rev. Lett. \textbf{95} (2005), 021601
[arXiv:hep-th/0403049 [hep-th]].

\bibitem{Derendinger:2004jn}
J.~P.~Derendinger, C.~Kounnas, P.~M.~Petropoulos and F.~Zwirner,
``Superpotentials in IIA compactifications with general fluxes,''
Nucl. Phys. B \textbf{715} (2005), 211-233
[arXiv:hep-th/0411276 [hep-th]].

\bibitem{Cribiori:2021djm}
N.~Cribiori, D.~Junghans, V.~Van Hemelryck, T.~Van Riet and T.~Wrase,
``Scale-separated AdS4 vacua of IIA orientifolds and M-theory,''
Phys. Rev. D \textbf{104} (2021) no.12, 126014
[arXiv:2107.00019 [hep-th]].

\bibitem{Farakos:2020phe}
F.~Farakos, G.~Tringas and T.~Van Riet,
``No-scale and scale-separated flux vacua from IIA on G2 orientifolds,''
Eur. Phys. J. C \textbf{80} (2020) no.7, 659
[arXiv:2005.05246 [hep-th]].

\bibitem{VanHemelryck:2022ynr}
V.~Van Hemelryck,
``Scale-Separated AdS3 Vacua from G2-Orientifolds Using Bispinors,''
Fortsch. Phys. \textbf{70} (2022) no.12, 2200128
[arXiv:2207.14311 [hep-th]].

\bibitem{Farakos:2023nms}
F.~Farakos, M.~Morittu and G.~Tringas,
``On/off scale separation,''
[arXiv:2304.14372 [hep-th]].

\bibitem{Junghans:2020acz}
D.~Junghans,
``O-Plane Backreaction and Scale Separation in Type IIA Flux Vacua,''
Fortsch. Phys. \textbf{68} (2020) no.6, 2000040
[arXiv:2003.06274 [hep-th]].


\bibitem{Marchesano:2020qvg}
F.~Marchesano, E.~Palti, J.~Quirant and A.~Tomasiello,
``On supersymmetric AdS$_{4}$ orientifold vacua,''
JHEP \textbf{08} (2020), 087
[arXiv:2003.13578 [hep-th]].

\bibitem{Emelin:2022cac}
M.~Emelin, F.~Farakos and G.~Tringas,
``O6-plane backreaction on scale-separated Type IIA AdS$_{3}$ vacua,''
JHEP \textbf{07} (2022), 133
[arXiv:2202.13431 [hep-th]].

\bibitem{Hristov:2009uj}
K.~Hristov, H.~Looyestijn and S.~Vandoren,
``Maximally supersymmetric solutions of D=4 N=2 gauged supergravity,''
JHEP \textbf{11} (2009), 115
[arXiv:0909.1743 [hep-th]].

\bibitem{Lust:2017aqj}
S.~Lust, P.~Ruter and J.~Louis,
``Maximally Supersymmetric AdS Solutions and their Moduli Spaces,''
JHEP \textbf{03} (2018), 019
[arXiv:1711.06180 [hep-th]].

\bibitem{Cribiori:2020use}
N.~Cribiori, G.~Dall'agata and F.~Farakos,
``Weak gravity versus de Sitter,''
JHEP \textbf{04} (2021), 046
[arXiv:2011.06597 [hep-th]].

\bibitem{Montero:2022ghl}
M.~Montero, M.~Rocek and C.~Vafa,
``Pure supersymmetric AdS and the Swampland,''
JHEP \textbf{01} (2023), 094
[arXiv:2212.01697 [hep-th]].

\bibitem{Cribiori:2023ihv}
N.~Cribiori and C.~Montella,
``Quantum gravity constraints on scale separation and de Sitter in five dimensions,''
[arXiv:2303.04162 [hep-th]].

\bibitem{DallAgata:2021nnr}
G.~Dall'Agata, M.~Emelin, F.~Farakos and M.~Morittu,
``The unbearable lightness of charged gravitini,''
JHEP \textbf{10} (2021), 076
[arXiv:2108.04254 [hep-th]].

\bibitem{Emelin:2022wft}
M.~Emelin,
``Obstacles for dS in Supersymmetric Theories,''
PoS \textbf{CORFU2021} (2022), 187
[arXiv:2206.01603 [hep-th]].

\bibitem{Cribiori:2022sxf}
N.~Cribiori,
``De Sitter, gravitino mass and the swampland,''
PoS \textbf{CORFU2021} (2022), 200
[arXiv:2203.15449 [hep-th]].

\bibitem{Kachru:2003aw}
S.~Kachru, R.~Kallosh, A.~D.~Linde and S.~P.~Trivedi,
``De Sitter vacua in string theory,''
Phys. Rev. D \textbf{68} (2003), 046005
[arXiv:hep-th/0301240 [hep-th]].

\bibitem{Balasubramanian:2005zx}
V.~Balasubramanian, P.~Berglund, J.~P.~Conlon and F.~Quevedo,
``Systematics of moduli stabilisation in Calabi-Yau flux compactifications,''
JHEP \textbf{03} (2005), 007
[arXiv:hep-th/0502058 [hep-th]].

\bibitem{Danielsson:2018ztv}
U.~H.~Danielsson and T.~Van Riet,
``What if string theory has no de Sitter vacua?,''
Int. J. Mod. Phys. D \textbf{27} (2018) no.12, 1830007
[arXiv:1804.01120 [hep-th]].

\bibitem{Gao:2020xqh}
X.~Gao, A.~Hebecker and D.~Junghans,
``Control issues of KKLT,''
Fortsch. Phys. \textbf{68} (2020), 2000089
[arXiv:2009.03914 [hep-th]].



\bibitem{Junghans:2022exo}
D.~Junghans,
``LVS de Sitter vacua are probably in the swampland,''
Nucl. Phys. B \textbf{990} (2023), 116179
[arXiv:2201.03572 [hep-th]].

\bibitem{Gao:2022fdi}
X.~Gao, A.~Hebecker, S.~Schreyer and G.~Venken,
``The LVS parametric tadpole constraint,''
JHEP \textbf{07} (2022), 056
[arXiv:2202.04087 [hep-th]].

\bibitem{Junghans:2022kxg}
D.~Junghans,
``Topological constraints in the LARGE-volume scenario,''
JHEP \textbf{08} (2022), 226
[arXiv:2205.02856 [hep-th]].

\bibitem{Blumenhagen:2022dbo}
R.~Blumenhagen, A.~Gligovic and S.~Kaddachi,
``Mass Hierarchies and Quantum Gravity Constraints in DKMM-refined KKLT,''
Fortsch. Phys. \textbf{71} (2023) no.1, 2200167
[arXiv:2206.08400 [hep-th]].

\bibitem{Obied:2018sgi}
G.~Obied, H.~Ooguri, L.~Spodyneiko and C.~Vafa,
``De Sitter Space and the Swampland,''
[arXiv:1806.08362 [hep-th]].

\bibitem{Garg:2018reu}
S.~K.~Garg and C.~Krishnan,
JHEP \textbf{11} (2019), 075
[arXiv:1807.05193 [hep-th]].

\bibitem{Andriot:2018wzk}
D.~Andriot,
``On the de Sitter swampland criterion,''
Phys. Lett. B \textbf{785} (2018), 570-573
[arXiv:1806.10999 [hep-th]].

\bibitem{Ooguri:2018wrx}
H.~Ooguri, E.~Palti, G.~Shiu and C.~Vafa,
``Distance and de Sitter Conjectures on the Swampland,''
Phys. Lett. B \textbf{788} (2019), 180-184
[arXiv:1810.05506 [hep-th]].

\bibitem{Bedroya:2019snp}
A.~Bedroya and C.~Vafa,
``Trans-Planckian Censorship and the Swampland,''
JHEP \textbf{09} (2020), 123
[arXiv:1909.11063 [hep-th]].

\bibitem{Cicoli:2021fsd}
M.~Cicoli, F.~Cunillera, A.~Padilla and F.~G.~Pedro,
``Quintessence and the Swampland: The Parametrically Controlled Regime of Moduli Space,''
Fortsch. Phys. \textbf{70} (2022) no.4, 2200009
[arXiv:2112.10779 [hep-th]].

\bibitem{Cicoli:2021skd}
M.~Cicoli, F.~Cunillera, A.~Padilla and F.~G.~Pedro,
``Quintessence and the Swampland: The Numerically Controlled Regime of Moduli Space,''
Fortsch. Phys. \textbf{70} (2022) no.4, 2200008
[arXiv:2112.10783 [hep-th]].


\bibitem{Ferrara:2019tmu}
S.~Ferrara, M.~Tournoy and A.~Van Proeyen,
``de Sitter Conjectures in $N$=1 Supergravity,''
Fortsch. Phys. \textbf{68} (2020) no.2, 1900107
[arXiv:1912.06626 [hep-th]].

\bibitem{Andriot:2021rdy}
D.~Andriot,
``Tachyonic de Sitter Solutions of 10d Type II Supergravities,''
Fortsch. Phys. \textbf{69} (2021) no.7, 2100063
[arXiv:2101.06251 [hep-th]].

\bibitem{Andriot:2022way}
D.~Andriot, L.~Horer and P.~Marconnet,
``Charting the landscape of (anti-) de Sitter and Minkowski solutions of 10d supergravities,''
JHEP \textbf{06} (2022), 131
[arXiv:2201.04152 [hep-th]].


\bibitem{Cribiori:2020wch}
N.~Cribiori, F.~Farakos and G.~Tringas,
``Three-forms and Fayet-Iliopoulos terms in Supergravity: Scanning Planck mass and BPS domain walls,''
JHEP \textbf{05} (2020), 060
[arXiv:2001.05757 [hep-th]].

\bibitem{Cribiori:2017laj}
N.~Cribiori, F.~Farakos, M.~Tournoy and A.~van Proeyen,
``Fayet-Iliopoulos terms in supergravity without gauged R-symmetry,''
JHEP \textbf{04} (2018), 032
[arXiv:1712.08601 [hep-th]].

\bibitem{Kuzenko:2018jlz}
S.~M.~Kuzenko,
``Taking a vector supermultiplet apart: Alternative Fayet\textendash{}Iliopoulos-type terms,''
Phys. Lett. B \textbf{781} (2018), 723-727
[arXiv:1801.04794 [hep-th]].


\bibitem{Antoniadis:2018cpq}
I.~Antoniadis, A.~Chatrabhuti, H.~Isono and R.~Knoops,
``Fayet\textendash{}Iliopoulos terms in supergravity and D-term inflation,''
Eur. Phys. J. C \textbf{78} (2018) no.5, 366
[arXiv:1803.03817 [hep-th]].

\bibitem{Antoniadis:2018oeh}
I.~Antoniadis, A.~Chatrabhuti, H.~Isono and R.~Knoops,
``The cosmological constant in Supergravity,''
Eur. Phys. J. C \textbf{78} (2018) no.9, 718
[arXiv:1805.00852 [hep-th]].

\bibitem{Antoniadis:2019hbu}
I.~Antoniadis, J.~P.~Derendinger, F.~Farakos and G.~Tartaglino-Mazzucchelli,
``New Fayet-Iliopoulos terms in $ \mathcal{N}=2 $ supergravity,''
JHEP \textbf{07} (2019), 061
[arXiv:1905.09125 [hep-th]].

\bibitem{Seiberg:2010qd}
N.~Seiberg,
``Modifying the Sum Over Topological Sectors and Constraints on Supergravity,''
JHEP \textbf{07} (2010), 070
[arXiv:1005.0002 [hep-th]].

\bibitem{Distler:2010zg}
J.~Distler and E.~Sharpe,
``Quantization of Fayet-Iliopoulos Parameters in Supergravity,''
Phys. Rev. D \textbf{83} (2011), 085010
[arXiv:1008.0419 [hep-th]].

\bibitem{Hellerman:2010fv}
S.~Hellerman and E.~Sharpe,
``Sums over topological sectors and quantization of Fayet-Iliopoulos parameters,''
Adv. Theor. Math. Phys. \textbf{15} (2011), 1141-1199
[arXiv:1012.5999 [hep-th]].


\bibitem{Komargodski:2009pc}
Z.~Komargodski and N.~Seiberg,
``Comments on the Fayet-Iliopoulos Term in Field Theory and Supergravity,''
JHEP \textbf{06} (2009), 007
[arXiv:0904.1159 [hep-th]].


\bibitem{Lauria:2020rhc}
E.~Lauria and A.~Van Proeyen,
``${\cal N}=2$ Supergravity in $D=4,5,6$ Dimensions,''
Lect. Notes Phys. \textbf{966} (2020), pp.
2020,
ISBN 978-3-030-33755-1, 978-3-030-33757-5
[arXiv:2004.11433 [hep-th]].


\bibitem{Fre:2002pd}
P.~Fre, M.~Trigiante and A.~Van Proeyen,
``Stable de Sitter vacua from N=2 supergravity,''
Class. Quant. Grav. \textbf{19} (2002), 4167-4194
[arXiv:hep-th/0205119 [hep-th]].

\bibitem{Ceresole:2001wi}
A.~Ceresole, G.~Dall'Agata, R.~Kallosh and A.~Van Proeyen,
``Hypermultiplets, domain walls and supersymmetric attractors,''
Phys. Rev. D \textbf{64} (2001), 104006
[arXiv:hep-th/0104056 [hep-th]].


\bibitem{Catino:2013syn}
F.~Catino, C.~A.~Scrucca and P.~Smyth,
``Simple metastable de Sitter vacua in N=2 gauged supergravity,''
JHEP \textbf{04} (2013), 056
[arXiv:1302.1754 [hep-th]].


\bibitem{Montero:2019ekk}
M.~Montero, T.~Van Riet and G.~Venken,
``Festina Lente: EFT Constraints from Charged Black Hole Evaporation in de Sitter,''
JHEP \textbf{01} (2020), 039
[arXiv:1910.01648 [hep-th]].



\bibitem{Montero:2021otb}
M.~Montero, C.~Vafa, T.~Van Riet and G.~Venken,
``The FL bound and its phenomenological implications,''
JHEP \textbf{10} (2021), 009
[arXiv:2106.07650 [hep-th]].



\bibitem{Lee:2021cor}
S.~M.~Lee, D.~Y.~Cheong, S.~C.~Hyun, S.~C.~Park and M.~S.~Seo,
``Festina-Lente bound on Higgs vacuum structure and inflation,''
JHEP \textbf{02} (2022), 100
[arXiv:2111.04010 [hep-ph]].


\bibitem{Ban:2022jgm}
K.~Ban, D.~Y.~Cheong, H.~Okada, H.~Otsuka, J.~C.~Park and S.~C.~Park,
``Phenomenological implications on a hidden sector from the festina lente bound,''
PTEP \textbf{2023} (2023) no.1, 013B04
[arXiv:2206.00890 [hep-ph]].


\bibitem{Guidetti:2022xct}
V.~Guidetti, N.~Righi, G.~Venken and A.~Westphal,
``Axionic Festina Lente,''
JHEP \textbf{01} (2023), 114
[arXiv:2206.03494 [hep-th]].

\bibitem{Mohseni:2022ftn}
A.~Mohseni and M.~Torabian,
``Higgs in Nilpotent Supergravity: Vacuum Energy and Festina Lente,''
[arXiv:2207.13593 [hep-th]].

\bibitem{Palti:2017elp}
E.~Palti,
``The Weak Gravity Conjecture and Scalar Fields,''
JHEP \textbf{08} (2017), 034
[arXiv:1705.04328 [hep-th]].


\bibitem{Lust:2017wrl}
D.~Lust and E.~Palti,
``Scalar Fields, Hierarchical UV/IR Mixing and The Weak Gravity Conjecture,''
JHEP \textbf{02} (2018), 040
[arXiv:1709.01790 [hep-th]].

\bibitem{Gonzalo:2019gjp}
E.~Gonzalo and L.~E.~Ib\'a\~nez,
``A Strong Scalar Weak Gravity Conjecture and Some Implications,''
JHEP \textbf{08} (2019), 118
[arXiv:1903.08878 [hep-th]].

\bibitem{Benakli:2020pkm}
K.~Benakli, C.~Branchina and G.~Lafforgue-Marmet,
``Revisiting the scalar weak gravity conjecture,''
Eur. Phys. J. C \textbf{80} (2020) no.8, 742
[arXiv:2004.12476 [hep-th]].

\bibitem{DallAgata:2020ino}
G.~Dall'Agata and M.~Morittu,
``Covariant formulation of BPS black holes and the scalar weak gravity conjecture,''
JHEP \textbf{03} (2020), 192
[arXiv:2001.10542 [hep-th]].


\bibitem{Polchinski:2003bq}
J.~Polchinski,
``Monopoles, duality, and string theory,''
Int. J. Mod. Phys. A \textbf{19S1} (2004), 145-156
[arXiv:hep-th/0304042 [hep-th]].


\bibitem{Lanza:2020qmt}
S.~Lanza, F.~Marchesano, L.~Martucci and I.~Valenzuela,
``Swampland Conjectures for Strings and Membranes,''
JHEP \textbf{02} (2021), 006
[arXiv:2006.15154 [hep-th]].


\bibitem{Lanza:2021udy}
S.~Lanza, F.~Marchesano, L.~Martucci and I.~Valenzuela,
``The EFT stringy viewpoint on large distances,''
JHEP \textbf{09} (2021), 197
[arXiv:2104.05726 [hep-th]].


\bibitem{Martucci:2022krl}
L.~Martucci, N.~Risso and T.~Weigand,
``Quantum gravity bounds on $ \mathcal{N} $ = 1 effective theories in four dimensions,''
JHEP \textbf{03} (2023), 197
[arXiv:2210.10797 [hep-th]].

\bibitem{Marchesano:2022avb}
F.~Marchesano and M.~Wiesner,
``4d strings at strong coupling,''
JHEP \textbf{08} (2022), 004
[arXiv:2202.10466 [hep-th]].


\bibitem{Marchesano:2022axe}
F.~Marchesano and L.~Melotti,
``EFT strings and emergence,''
JHEP \textbf{02} (2023), 112
[arXiv:2211.01409 [hep-th]].


\bibitem{Katz:2020ewz}
S.~Katz, H.~C.~Kim, H.~C.~Tarazi and C.~Vafa,
``Swampland Constraints on 5d $\mathcal{N}=1$ Supergravity,''
JHEP \textbf{07} (2020), 080
[arXiv:2004.14401 [hep-th]].

\bibitem{Farakos:2017jme}
F.~Farakos, S.~Lanza, L.~Martucci and D.~Sorokin,
``Three-forms in Supergravity and Flux Compactifications,''
Eur. Phys. J. C \textbf{77} (2017) no.9, 602
[arXiv:1706.09422 [hep-th]].


\bibitem{Bandos:2018gjp}
I.~Bandos, F.~Farakos, S.~Lanza, L.~Martucci and D.~Sorokin,
``Three-forms, dualities and membranes in four-dimensional supergravity,''
JHEP \textbf{07} (2018), 028
[arXiv:1803.01405 [hep-th]].


\bibitem{Binetruy:1996xw}
P.~Binetruy, F.~Pillon, G.~Girardi and R.~Grimm,
``The Three form multiplet in supergravity,''
Nucl. Phys. B \textbf{477} (1996), 175-202
[arXiv:hep-th/9603181 [hep-th]].

\bibitem{Ovrut:1997ur}
B.~A.~Ovrut and D.~Waldram,
``Membranes and three form supergravity,''
Nucl. Phys. B \textbf{506} (1997), 236-266
[arXiv:hep-th/9704045 [hep-th]].


\bibitem{Greene:1989ya}
B.~R.~Greene, A.~D.~Shapere, C.~Vafa and S.~T.~Yau,
``Stringy Cosmic Strings and Noncompact Calabi-Yau Manifolds,''
Nucl. Phys. B \textbf{337} (1990), 1-36
doi:10.1016/0550-3213(90)90248-C


\bibitem{Dabholkar:1990yf}
A.~Dabholkar, G.~W.~Gibbons, J.~A.~Harvey and F.~Ruiz Ruiz,
``Superstrings and Solitons,''
Nucl. Phys. B \textbf{340} (1990), 33-55

\bibitem{Gates:1980ay}
S.~J.~Gates, Jr.,
``SUPER P FORM GAUGE SUPERFIELDS,''
Nucl. Phys. B \textbf{184} (1981), 381-390

\bibitem{Binetruy:2000zx}
P.~Binetruy, G.~Girardi and R.~Grimm,
``Supergravity couplings: A Geometric formulation,''
Phys. Rept. \textbf{343} (2001), 255-462
[arXiv:hep-th/0005225 [hep-th]].

\bibitem{Cribiori:2018jjh}
N.~Cribiori and S.~Lanza,
``On the dynamical origin of parameters in $\mathcal {N}=2$ supersymmetry,''
Eur. Phys. J. C \textbf{79} (2019) no.1, 32
[arXiv:1810.11425 [hep-th]].


\bibitem{Kaloper:2022oqv}
N.~Kaloper,
``Hidden variables of gravity and geometry and the cosmological constant problem,''
Phys. Rev. D \textbf{106} (2022) no.6, 6
[arXiv:2202.06977 [hep-th]].


\bibitem{Kaloper:2022utc}
N.~Kaloper,
``Pancosmic Relativity and Nature's Hierarchies,''
Phys. Rev. D \textbf{106} (2022), 4
[arXiv:2202.08860 [hep-th]].


\bibitem{Kaloper:2022jpv}
N.~Kaloper and A.~Westphal,
``Quantum-mechanical mechanism for reducing the cosmological constant,''
Phys. Rev. D \textbf{106} (2022) no.10, L101701
[arXiv:2204.13124 [hep-th]].

\bibitem{Nakahara:2003nw}
M.~Nakahara,
``Geometry, topology and physics,''


\bibitem{Witten:1996hc}
E.~Witten,
``Five-brane effective action in M theory,''
J. Geom. Phys. \textbf{22} (1997), 103-133
[arXiv:hep-th/9610234 [hep-th]].

\bibitem{BottTu}
R.~Bott and W.~Tu, L., ``Differential Forms in Algebraic Topology,'' Springer New York, NY, 1982.



\bibitem{Alvarez-Gaume:1983ihn}
L.~Alvarez-Gaume and E.~Witten,
``Gravitational Anomalies,''
Nucl. Phys. B \textbf{234} (1984), 269


\bibitem{Farakos:2020wfc}
F.~Farakos, A.~Kehagias and N.~Liatsos,
``de Sitter decay through goldstino evaporation,''
JHEP \textbf{02} (2021), 186
[arXiv:2009.03335 [hep-th]].



\bibitem{Volkov:1973ix}
D.~V.~Volkov and V.~P.~Akulov,
``Is the Neutrino a Goldstone Particle?,''
Phys. Lett. B \textbf{46} (1973), 109-110


\bibitem{Lindstrom:1979kq}
U.~Lindstrom and M.~Rocek,
``CONSTRAINED LOCAL SUPERFIELDS,''
Phys. Rev. D \textbf{19} (1979), 2300-2303


\bibitem{Dudas:2015eha}
E.~Dudas, S.~Ferrara, A.~Kehagias and A.~Sagnotti,
``Properties of Nilpotent Supergravity,''
JHEP \textbf{09} (2015), 217
[arXiv:1507.07842 [hep-th]].


\bibitem{Bergshoeff:2015tra}
E.~A.~Bergshoeff, D.~Z.~Freedman, R.~Kallosh and A.~Van Proeyen,
``Pure de Sitter Supergravity,''
Phys. Rev. D \textbf{92} (2015) no.8, 085040
[erratum: Phys. Rev. D \textbf{93} (2016) no.6, 069901]
[arXiv:1507.08264 [hep-th]].

\bibitem{DallAgata:2016syy}
G.~Dall'Agata, E.~Dudas and F.~Farakos,
``On the origin of constrained superfields,''
JHEP \textbf{05} (2016), 041
[arXiv:1603.03416 [hep-th]].

\bibitem{Cribiori:2016qif}
N.~Cribiori, G.~Dall'Agata, F.~Farakos and M.~Porrati,
``Minimal Constrained Supergravity,''
Phys. Lett. B \textbf{764} (2017), 228-232
[arXiv:1611.01490 [hep-th]].

\bibitem{Farakos:2016hly}
F.~Farakos, A.~Kehagias, D.~Racco and A.~Riotto,
``Scanning of the Supersymmetry Breaking Scale and the Gravitino Mass in Supergravity,''
JHEP \textbf{06} (2016), 120
[arXiv:1605.07631 [hep-th]].


\bibitem{DallAgata:2022abm}
G.~Dall'Agata, M.~Emelin, F.~Farakos and M.~Morittu,
``Anti-brane uplift instability from goldstino condensation,''
JHEP \textbf{08} (2022), 005
[arXiv:2203.12636 [hep-th]].

\bibitem{Kallosh:2022fsc}
R.~Kallosh, A.~Linde, T.~Wrase and Y.~Yamada,
``Goldstino condensation?,''
JHEP \textbf{08} (2022), 166
[arXiv:2206.04210 [hep-th]].

\bibitem{Farakos:2022jcl}
F.~Farakos and M.~Morittu,
``Goldstino condensation at large N,''
Eur. Phys. J. C \textbf{83} (2023) no.2, 166
[arXiv:2211.12527 [hep-th]].


\bibitem{Alexandre:2013iva}
J.~Alexandre, N.~Houston and N.~E.~Mavromatos,
``Dynamical Supergravity Breaking via the Super-Higgs Effect Revisited,''
Phys. Rev. D \textbf{88} (2013), 125017
[arXiv:1310.4122 [hep-th]].



\bibitem{Alexandre:2014lla}
J.~Alexandre, N.~Houston and N.~E.~Mavromatos,
``Inflation via Gravitino Condensation in Dynamically Broken Supergravity,''
Int. J. Mod. Phys. D \textbf{24} (2015) no.04, 1541004
[arXiv:1409.3183 [gr-qc]].


\bibitem{Cribiori:2021gbf}
N.~Cribiori, D.~Lust and M.~Scalisi,
``The gravitino and the swampland,''
JHEP \textbf{06} (2021), 071
[arXiv:2104.08288 [hep-th]].


\bibitem{Castellano:2021yye}
A.~Castellano, A.~Font, A.~Herraez and L.~E.~Ib\'a\~nez,
``A gravitino distance conjecture,''
JHEP \textbf{08} (2021), 092
[arXiv:2104.10181 [hep-th]].


\bibitem{Andriot:2022brg}
D.~Andriot, L.~Horer and G.~Tringas,
``Negative scalar potentials and the swampland: an Anti-Trans-Planckian Censorship Conjecture,''
[arXiv:2212.04517 [hep-th]].


\bibitem{Anchordoqui:2023oqm}
L.~A.~Anchordoqui, I.~Antoniadis, N.~Cribiori, D.~Lust and M.~Scalisi,
``The Scale of Supersymmetry Breaking and the Dark Dimension,''
[arXiv:2301.07719 [hep-th]].

\bibitem{Montero:2022prj}
M.~Montero, C.~Vafa and I.~Valenzuela,
``The dark dimension and the Swampland,''
JHEP \textbf{02} (2023), 022
[arXiv:2205.12293 [hep-th]].


\bibitem{Coudarchet:2021qwc}
T.~Coudarchet, E.~Dudas and H.~Partouche,
``Geometry of orientifold vacua and supersymmetry breaking,''
JHEP \textbf{07} (2021), 104
[arXiv:2105.06913 [hep-th]].

\bibitem{Castellano:2022bvr}
A.~Castellano, A.~Herr\'aez and L.~E.~Ib\'a\~nez,
``The emergence proposal in quantum gravity and the species scale,''
JHEP \textbf{06} (2023), 047
[arXiv:2212.03908 [hep-th]].


\bibitem{Palti:2020tsy}
E.~Palti,
``Fermions and the Swampland,''
Phys. Lett. B \textbf{808} (2020), 135617
[arXiv:2005.08538 [hep-th]].

\bibitem{Linde:1990flp}
A.~D.~Linde,
``Particle physics and inflationary cosmology,''
Contemp. Concepts Phys. \textbf{5} (1990), 1-362
[arXiv:hep-th/0503203 [hep-th]].


\bibitem{Lyth:1998xn}
D.~H.~Lyth and A.~Riotto,
``Particle physics models of inflation and the cosmological density perturbation,''
Phys. Rept. \textbf{314} (1999), 1-146
[arXiv:hep-ph/9807278 [hep-ph]].

\bibitem{Ferrara:2013rsa}
S.~Ferrara, R.~Kallosh, A.~Linde and M.~Porrati,
``Minimal Supergravity Models of Inflation,''
Phys. Rev. D \textbf{88} (2013) no.8, 085038
[arXiv:1307.7696 [hep-th]].


\bibitem{Farakos:2013cqa}
F.~Farakos, A.~Kehagias and A.~Riotto,
``On the Starobinsky Model of Inflation from Supergravity,''
Nucl. Phys. B \textbf{876} (2013), 187-200
[arXiv:1307.1137 [hep-th]].


\bibitem{Kallosh:2010xz}
R.~Kallosh, A.~Linde and T.~Rube,
``General inflaton potentials in supergravity,''
Phys. Rev. D \textbf{83} (2011), 043507
[arXiv:1011.5945 [hep-th]].


\bibitem{Kawasaki:2000yn}
M.~Kawasaki, M.~Yamaguchi and T.~Yanagida,
``Natural chaotic inflation in supergravity,''
Phys. Rev. Lett. \textbf{85} (2000), 3572-3575
[arXiv:hep-ph/0004243 [hep-ph]].


\bibitem{Goncharov:1983mw}
A.~B.~Goncharov and A.~D.~Linde,
``Chaotic Inflation in Supergravity,''
Phys. Lett. B \textbf{139} (1984), 27-30

\bibitem{Ceresole:2014vpa}
A.~Ceresole, G.~Dall'Agata, S.~Ferrara, M.~Trigiante and A.~Van Proeyen,
``A search for an $\mathcal{N} =2 $ inflaton potential,''
Fortsch. Phys. \textbf{62} (2014), 584-606
[arXiv:1404.1745 [hep-th]].


\bibitem{Dalianis:2014sqa}
I.~Dalianis and F.~Farakos,
``Higher Derivative D-term Inflation in New-minimal Supergravity,''
Phys. Lett. B \textbf{736} (2014), 299-304
[arXiv:1403.3053 [hep-th]].



\bibitem{DallAgata:2018ybl}
G.~Dall'Agata,
``Chromo-Natural inflation in Supergravity,''
Phys. Lett. B \textbf{782} (2018), 139-142
[arXiv:1804.03104 [hep-th]].


\bibitem{DallAgata:2019yrr}
G.~Dall'Agata, S.~Gonz\'alez-Mart\'\i{}n, A.~Papageorgiou and M.~Peloso,
``Warm dark energy,''
JCAP \textbf{08} (2020), 032
[arXiv:1912.09950 [hep-th]].

\bibitem{Scalisi:2018eaz}
M.~Scalisi and I.~Valenzuela,
``Swampland distance conjecture, inflation and $\alpha$-attractors,''
JHEP \textbf{08} (2019), 160
[arXiv:1812.07558 [hep-th]].

\bibitem{Dalianis:2014aya}
I.~Dalianis, F.~Farakos, A.~Kehagias, A.~Riotto and R.~von Unge,
``Supersymmetry Breaking and Inflation from Higher Curvature Supergravity,''
JHEP \textbf{01} (2015), 043
[arXiv:1409.8299 [hep-th]].

\bibitem{Planck:2018jri}
Y.~Akrami \textit{et al.} [Planck],
``Planck 2018 results. X. Constraints on inflation,''
Astron. Astrophys. \textbf{641} (2020), A10
[arXiv:1807.06211 [astro-ph.CO]].

\bibitem{Goldwirth:1991rj}
D.~S.~Goldwirth and T.~Piran,
``Initial conditions for inflation,''
Phys. Rept. \textbf{214} (1992), 223-291


\bibitem{Ijjas:2013vea}
A.~Ijjas, P.~J.~Steinhardt and A.~Loeb,
``Inflationary paradigm in trouble after Planck2013,''
Phys. Lett. B \textbf{723} (2013), 261-266
[arXiv:1304.2785 [astro-ph.CO]].


\bibitem{Dalianis:2015fpa}
I.~Dalianis and F.~Farakos,
``On the initial conditions for inflation with plateau potentials: the $R+R^2$ (super)gravity case,''
JCAP \textbf{07} (2015), 044
[arXiv:1502.01246 [gr-qc]].

\bibitem{Linde:2017pwt}
A.~Linde,
``On the problem of initial conditions for inflation,''
Found. Phys. \textbf{48} (2018) no.10, 1246-1260
[arXiv:1710.04278 [hep-th]].

\bibitem{Ferrara:2013kca}
S.~Ferrara, R.~Kallosh, A.~Linde and M.~Porrati,
``Higher Order Corrections in Minimal Supergravity Models of Inflation,''
JCAP \textbf{11} (2013), 046
[arXiv:1309.1085 [hep-th]].

\bibitem{VanProeyen:1979ks}
A.~Van Proeyen,
``Massive Vector Multiplets in Supergravity,''
Nucl. Phys. B \textbf{162} (1980), 376

\bibitem{Cecotti:1987qr}
S.~Cecotti, S.~Ferrara and L.~Girardello,
``Massive Vector Multiplets From Superstrings,''
Nucl. Phys. B \textbf{294} (1987), 537-555

\bibitem{Cecotti:1987qe}
S.~Cecotti, S.~Ferrara, M.~Porrati and S.~Sabharwal,
``NEW MINIMAL HIGHER DERIVATIVE SUPERGRAVITY COUPLED TO MATTER,''
Nucl. Phys. B \textbf{306} (1988), 160-180

\bibitem{Kallosh:2013yoa}
R.~Kallosh, A.~Linde and D.~Roest,
``Superconformal Inflationary $\alpha$-Attractors,''
JHEP \textbf{11} (2013), 198
[arXiv:1311.0472 [hep-th]].

\bibitem{Stewart:1994ts}
E.~D.~Stewart,
``Inflation, supergravity and superstrings,''
Phys. Rev. D \textbf{51} (1995), 6847-6853
[arXiv:hep-ph/9405389 [hep-ph]].


\end{thebibliography}
\end{document}